\documentclass[amsmath,amssymb,amsbsy,reprint,prb,preprintnumbers,showpacs]{revtex4-2}
\usepackage{graphicx,color}
\usepackage{dcolumn}
\usepackage{bm}
\usepackage{braket}
\usepackage{mathtools}
\usepackage{ulem}
\usepackage[breaklinks,colorlinks=true,linkcolor=blue,urlcolor=blue,citecolor=blue]{hyperref}
\usepackage{times}

\begin{document}
\title{Unconventional gapless semiconductor in an extended martini lattice in covalent honeycomb materials}
\author{Tomonari Mizoguchi}
\affiliation{
Department of Physics, Graduate School of Pure and Applied Sciences, University of Tsukuba, 1-1-1 Tennodai, Tsukuba, Ibaraki 305-8571, Japan
}
\email{mizoguchi@rhodia.ph.tsukuba.ac.jp}
\author{Yanlin Gao}
\affiliation{
Department of Physics, Graduate School of Pure and Applied Sciences, University of Tsukuba, 1-1-1 Tennodai, Tsukuba, Ibaraki 305-8571, Japan
}
\author{Mina Maruyama}
\affiliation{
Department of Physics, Graduate School of Pure and Applied Sciences, University of Tsukuba, 1-1-1 Tennodai, Tsukuba, Ibaraki 305-8571, Japan
}
\author{Yasuhiro Hatsugai}
\affiliation{
Department of Physics, Graduate School of Pure and Applied Sciences, University of Tsukuba, 1-1-1 Tennodai, Tsukuba, Ibaraki 305-8571, Japan
}
\author{Susumu Okada}
\affiliation{
Department of Physics, Graduate School of Pure and Applied Sciences, University of Tsukuba, 1-1-1 Tennodai, Tsukuba, Ibaraki 305-8571, Japan
}
\date{\today}

\begin{abstract}
We study characteristic electronic structures in an extended martini lattice model 
and propose its materialization in $\pi$-electron networks
constructed by designated chemisorption on graphene and silicene. 
By investigating the minimal tight-binding model, we reveal rich electronic structures tuned by the ratio of hopping parameters,
ranging from the band insulator to the unconventional gapless semiconductor. 
Remarkably, the unconventional gapless semiconductor is characterized by a flat band at the Fermi level.
Further, the density functional theory calculations for candidate materials reveal that 
the characteristic electronic structures can be realized by designated chemisorption or chemical substitution on graphene and silicene,
and that the electronic structure near the Fermi level is tunable by the choice of the atomic species of adsorbed atoms.
Our results open the way to search exotic electronic structures 
and their functionalities induced by an extended martini lattice.
\end{abstract}

\maketitle
\textit{Introduction.} Exotic electronic structures are a source of rich phenomena in solid-state physics. 
In particular, the density of states (DOS) near the Fermi energy is one of the key quantities for
the determination of physical properties, 
such as phase transitions and the response to external electric and magnetic fields. 
In this regard, bands with constant energy in the entire Brillouin zone, called flat bands, 
are of particular interest because they provide diverging DOS, which implies instability.
Indeed, when flat bands are present near the Fermi energy,
various correlation-induced phases such as ferromagnetism~\cite{Mielke1991,Tasaki1992} and superconductivity~\cite{Imada2000,Heikkila2011,Peotta2015,Aoki2020,Peri2021} 
are predicted to be realized.
Also it may induce a structural deformation (i.e., the Peierls instability).

Recently, the search for flat-band materials has become active~\cite{Calugaru2022,Regnault2022,Chiu2022}.
Among various routes for realizing flat bands, 
electronic structure engineering in two-dimensional materials with $\pi$-electron networks,
such as graphene~\cite{Wallace1947,Novoselov2004,Novoselov2005,Geim2007,CastroNeto2009} and silicene~\cite{Garcia2011,Vogt2012,Fleurence2012}, has attracted considerable attention. 
One of the most prominent findings along this line is the emergence of strongly correlated physics and superconductivity 
in twisted bilayer graphene~\cite{Cao2018,Cao2018_2,Yankowitz2019,Park2021},
which originates from twist-induced flat bands, or the moir\'{e} flat bands,
appearing in designated twist angles called magic angles~\cite{Tarnopolsky2019}.
Another promising method for electronic structure engineering is 
fabricating superstructures~\cite{Shima1993} by making holes or chemical substitutions/adsorptions.
$\pi$ electrons do not have in-plane anisotropy, so their electronic structures are largely 
dominated by geometrical structures of lattices they live on.
This fact leads to a clear guiding principle for realizing flat bands, namely, 
to arrange the $\pi$-electron network such that it is equivalent to the famous flat-band lattice models.
For instance, the flat bands originating from a kagome-like network appear 
in graphene with periodic holes~\cite{Maruyama2016,Maruyama2017} 
and hydrocarbon networks containing $sp^2$ and $sp^3$ carbons~\cite{Sorimachi2017,Fujii2018,Fujii2018_2}
(i.e., the covalent organic framework).

\begin{figure}[b]
\begin{center}
\includegraphics[width=0.95\linewidth]{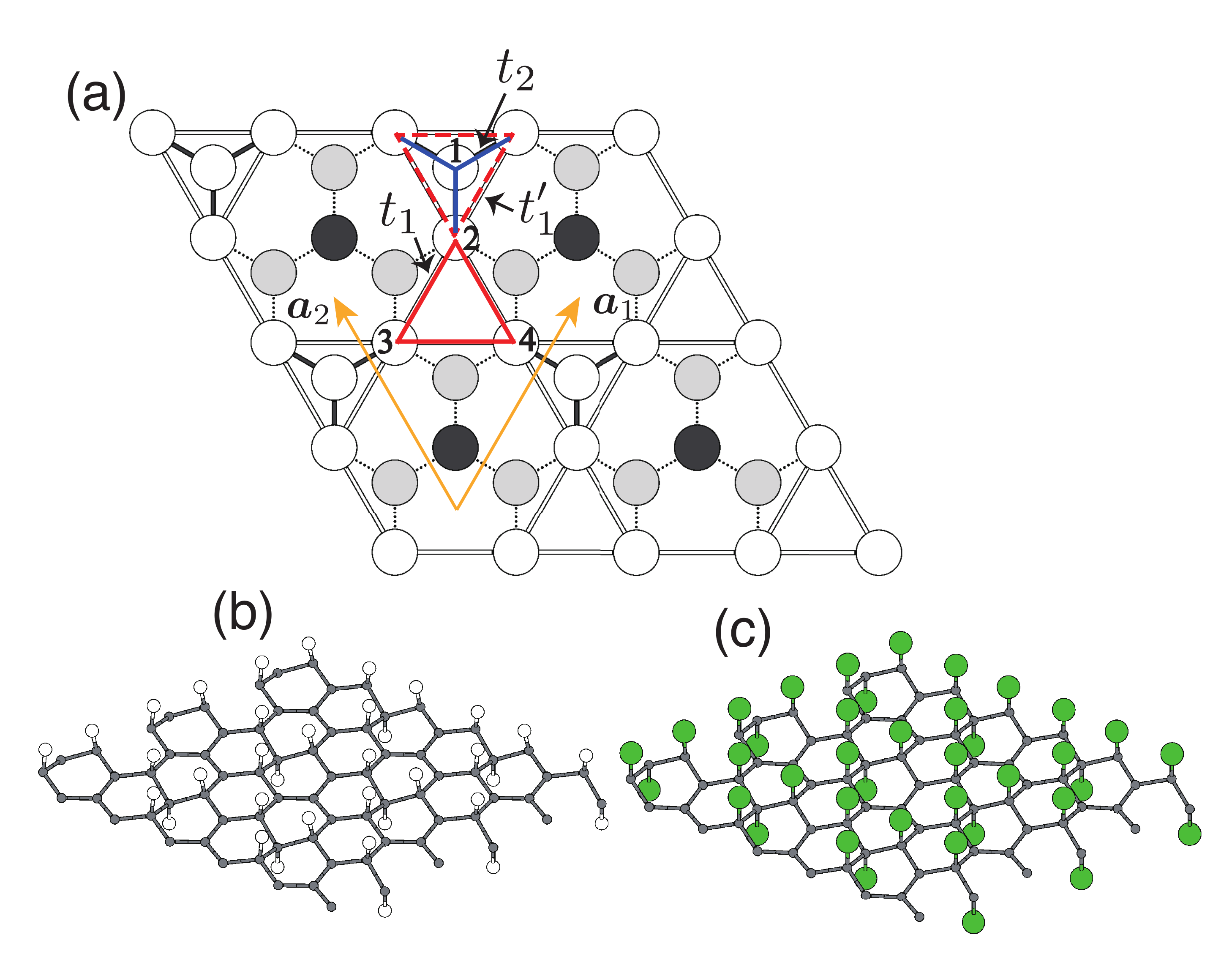}
\caption{
(a) Schematic views of the extended martini networks derived 
from hexagonal covalent networks of C and Si.
White and gray circles denote the C and Si 
without and with adsorbates, respectively. 
The adsorbates are adsorbed on gray atomic sites in (a) from the top and bottom of the layer.
Orange arrows represent the lattice vectors: 
$\bm{a}_1 = a_0 \left( \frac{1}{2}, \frac{\sqrt{3}}{2} \right)$ and $\bm{a}_2 = a_0 \left( -\frac{1}{2}, \frac{\sqrt{3}}{2} \right)$, 
where $a_0$ is the length of the unit cell edge.
The on-site potential $V$ is introduced on sublattice 1. 
Optimized geometric structures of (b) partially hydrogenated graphene, 
and (c) partially fluorinated 
graphene. 
Gray, green, and white spheres denote C, F, and H atoms, respectively.
}
\label{geometry}
\end{center}
\end{figure}
In this Letter, we propose a characteristic $\pi$-electron network hosting a flat band, 
which we refer to as an extended martini lattice [Fig.~\ref{geometry}(a)].
The network consists of corner sharing triangles (solid and dashed red bonds), 
which is equivalent to the kagome lattice, and the Y-shaped units (blue bonds) centered at the downward triangles.
Without downward triangles, (i.e., the dashed red bonds), the lattice is called a martini lattice~\cite{Miyahara2005,Kubo2006,Scullard2006,McClarty2020,Matsumoto2022},
which belongs to a class of flat-band lattices called a partial line graph~\cite{Miyahara2005,Kubo2006}.
Hence, the extension we consider here refers to the existence of the downward triangles.  
\begin{figure*}[t]
\begin{center}
\includegraphics[clip,width = 0.9\linewidth]{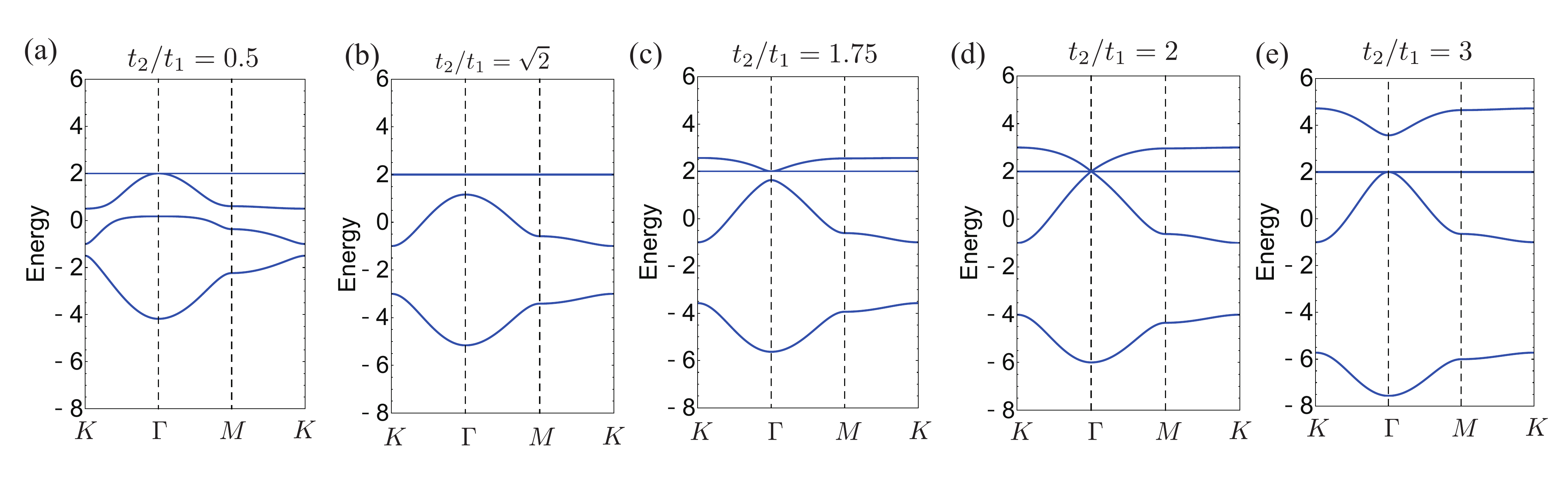}
\vspace{-10pt}
\caption{Band structure the for a tight-binding model along the high-symmetry lines in the Brillouin zone.
The high-symmetry points are $\Gamma = (0,0)$, $K = \left(\frac{4\pi}{3 a_0}, 0 \right)$, and $M = \left(\frac{\pi}{a_0}, \frac{\sqrt{3} \pi}{3a_0}\right)$.
We fix $t_1  =t_1^\prime = -1$, $V = 0$ and vary $t_2 /t_1$ whose value is shown at the top of each panel.
}
  \label{fig:band}
 \end{center}
 \vspace{-10pt}
\end{figure*}

We first elucidate the characteristic band structures in the extended martini lattice by employing a minimal tight-binding model.  
We find that the electronic structure strongly depends on the ratio between two hopping parameters. 
In particular, when the ratio exceeds a critical value, the lattice system becomes an unconventional gapless semiconductor where
a flat band is at the conduction band edge or the valence band edge.
We further show, based on density functional theory, 
that an extended martini lattice can be realized by partial chemisorption of graphene and silicene. 
In those materials, the flat bands acquire a finite dispersion, resulting in carrier doping to the flat bands accompanied by ferromagnetic ordering. 
Remarkably, the electronic structure near the band edge can be controlled by species 
of adsorbed atoms, which will open a way to search for exotic properties and functionalities in these systems. 

\textit{Robust flat band in an extended martini lattice model.}
The main target of this Letter is the extended martini lattice model, shown in Fig.~\ref{geometry}(a).
There are four sublattices per unit cell. 
Thus, in a tight-binding model for spinless, single-orbital fermions, the Bloch Hamiltonian $ H_{\bm{k}}$ has the form of 
a $4 \times 4$ matrix. The explicit form of $H_{\bm{k}}$ is shown in the Supplemental Material~\cite{SM}.
There are four parameters: $t_1$, $t_1^\prime$, $t_2$, and $V$.
Note that a conventional martini lattice model, which belongs to the partial line graph~\cite{Miyahara2005}, 
corresponds to the case of $t_1^\prime =V= 0$.

We elucidate that an exact flat band appears for any parameters, using the wisdom of linear algebra~\cite{Hatsugai2011,Hatsugai2015,Mizoguchi2019,Mizoguchi2019_star,Mizoguchi2020,Mizoguchi2021_skagome}. 
To begin with, we introduce three column vectors:
$\bm{\psi}_{\bm{k},1} = (0,1,1,1)^{\rm T}$ and $\bm{\psi}_{\bm{k},2} =(0, 1, e^{-i \bm{k} \cdot \bm{a}_1}, e^{-i \bm{k} \cdot \bm{a}_2})^{\rm T}$, and $\bm{\psi}_{\bm{k},3} = (1,0,0,0)^{\rm T}$~\cite{Remark1}.
We additionally introduce a $4 \times 3$ matrix, 
$\Psi_{\bm{k}} = \left( \bm{\psi}_{\bm{k},1} \hspace{1mm} \bm{\psi}_{\bm{k},2}\hspace{1mm} \bm{\psi}_{\bm{k},3} \right)$.
Its Hermitian conjugate, $\Psi_{\bm{k}}^\dagger$, is the $3 \times 4$ matrix. 
It follows that, for any $\bm{k}$, there exists a four-component
vector $\bm{\varphi}_{\bm{k}}$ that satisfies $\Psi^\dagger_{\bm{k}} \bm{\varphi}_{\bm{k}} = \bm{0}$.
In other words, $\bm{\varphi}_{\bm{k}}$ belongs to the kernel of the linear map represented by $\Psi^\dagger_{\bm{k}}$.
Its explicit form can be easily obtained for generic $\bm{k}$,
\begin{eqnarray}
\bm{\varphi}_{\bm{k}} = \frac{1}{\mathcal{N}_{\bm{k}}} 
 \left(0,e^{i \bm{k} \cdot \bm{a}_2} -e^{i \bm{k} \cdot \bm{a}_1}, 1 - e^{i \bm{k} \cdot
\bm{a}_2} ,
e^{i \bm{k} \cdot \bm{a}_1}  -1\right)^{\mathrm{T}}, \nonumber \\ \label{eq:fbwf}
\end{eqnarray}
with $\mathcal{N}_{\bm{k}}$ being the normalization constant. 
Note that $\bm{\varphi}_{\bm{k}}$ has a vanishing amplitude at sublattice 1. 
The remaining finite components on sublattices 2-4 are identical to those of the flat band wave functions 
of a kagome lattice. 
Note also that $\bm{\varphi}_{\bm{k}} $ becomes a zero vector at $\bm{k} = (0,0)$ (i.e., $\Gamma$ point), namely, 
$\bm{\varphi}_{\bm{k}}$ is singular at this point.
We will address the implication of this fact later. 
\begin{figure}[b]
\begin{center}
\includegraphics[clip,width = 0.9\linewidth]{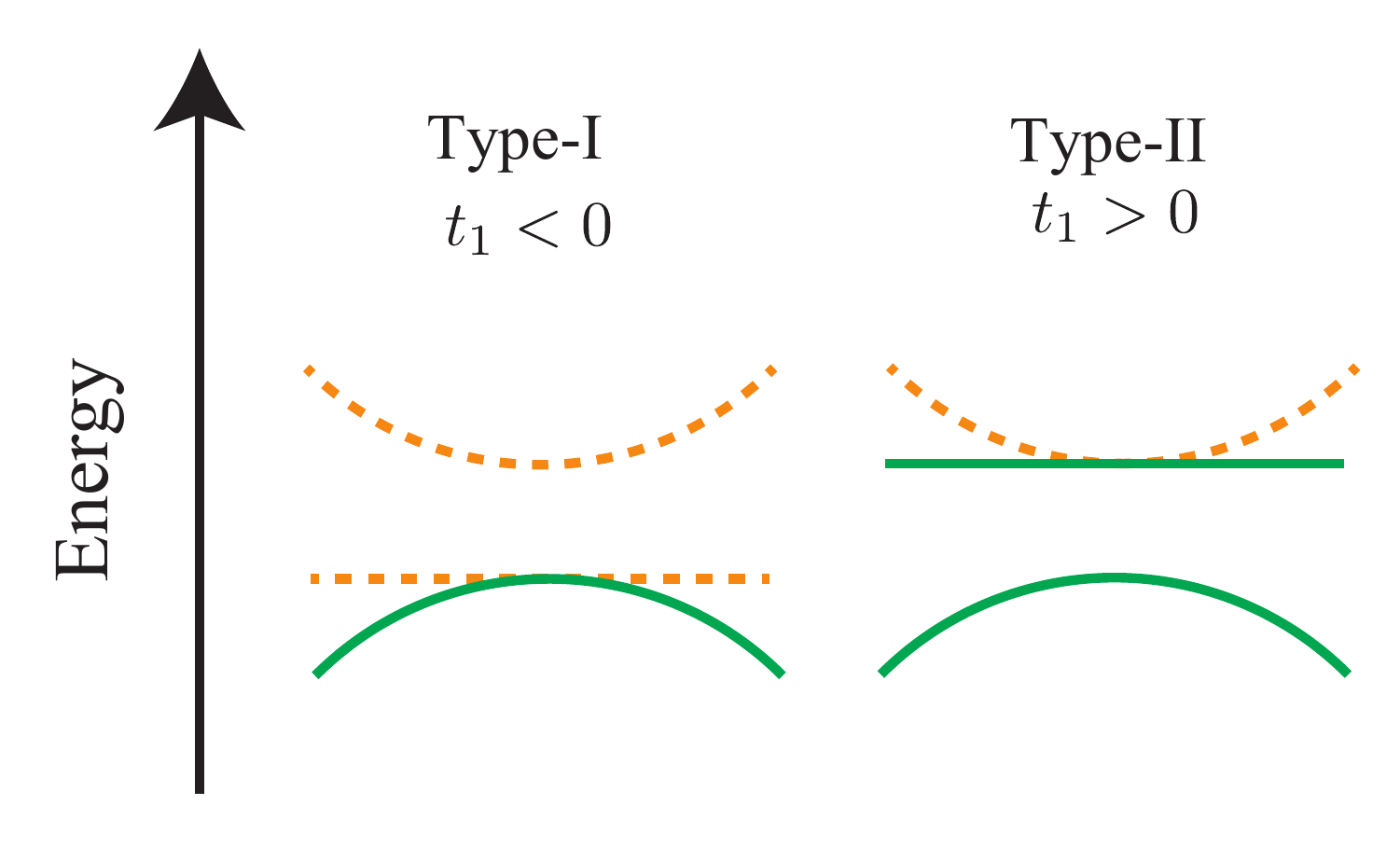}
\vspace{-10pt}
\caption{Schematic figure of the band structure around the Fermi levels for $|t_2| \gg |t_1|$
for $t_1 <0$ (i.e., type-I) and $t_1 > 0$ (i.e., type-II).
The solid green (dashed orange) bands are filled (empty).}
  \label{fig:types}
 \end{center}
 \vspace{-10pt}
\end{figure}

A key property for obtaining the flat band is that $H_{\bm{k}}$ 
can be expressed by the matrices introduced above:
\begin{eqnarray}
 H_{\bm{k}} = 
 \Psi_{\bm{k}} 
 \begin{pmatrix}
  t_1 & 0 & 0 \\
  0& t_1^\prime & t_2 \\
  0 & t_2 & V- \varepsilon^{\rm FB} \\
 \end{pmatrix}
 \Psi^\dagger_{\bm{k}} 
 + \varepsilon^{\rm FB} I_4, \label{eq:ham_exmartini_mo}
\end{eqnarray}
where $\varepsilon^{\rm FB} = - (t_1+t_1^\prime)$. 
Recalling that $\Psi^\dagger_{\bm{k}} \bm{\varphi}_{\bm{k}} = \bm{0}$ holds, 
we find $\bm{\varphi}_{\bm{k}}$ is the eigenvector of 
$H_{\bm{k}}$ with the eigenenergy being a $\bm{k}$-independent value, $\varepsilon^{\rm FB}$. 

\textit{Band structures of a tight-binding model.} 
We discuss the characteristics of the entire band structure, including an exact flat bands. 
In what follows, we focus on the case where $t_1= t_1^\prime = -1$ and $V = 0$, 
leaving an analysis of the generic parameters to the Supplemental Material~\cite{SM}.
In Fig.~\ref{fig:band}, we plot the band structures for several values of $t_2/t_1$. 
We see that, in all panels, 
an exact flat band with the energy being $\varepsilon^{\rm FB}$ indeed exists.
At the $\Gamma$ point, quadratic band touching between the flat band and the dispersive band occurs.
Such a band touching is ubiquitous in various flat band models~\cite{Bergman2008,Bilitewski2018,Rhim2019,Mizoguchi2019,Hwang2021,Hwang2021_2,Graf2021}. 
The band touching can be accounted for by the singularity of $\bm{\varphi_{\bm{k}}}$ which we have addressed before.
In fact, at $\Gamma$ point, 
$\bm{\psi}_{\bm{k},1} = \bm{\psi}_{\bm{k},2}$ 
holds, which results in the increase of the dimension of the kernel of $\Psi^\dagger_{\bm{k}}$ from 1 to 2.
Hence, only at this specific point, 
there are two eigenstates with the eigenenergy $\varepsilon= \varepsilon^{\rm FB}$, resulting in the quadratic band touching~\cite{Mizoguchi2019,Mizoguchi2020,Mizoguchi2021_skagome,Hatsugai2021,Mizoguchi2022,Kuroda2022}.

Besides these parameter-independent features, we also see in Fig.~\ref{fig:band} that varying $t_2/t_1$ causes the change of entire band structures.
Since the target materials are half-filled, we will focus on this case in the following discussions. 
For $0< t_2/t_1 < 2$ [Fig.~\ref{fig:band}(a)-(c)], the system is a conventional band insulator, 
where two dispersive bands are completely filled.
Interestingly, at fine-tuned parameter corresponding to Fig.~\ref{fig:band}(b), the flat band is doubly degenerate.
At $t_2/t_1 = 2$ [Fig.~\ref{fig:band}(d)], the triple band touching occurs at $\Gamma$ point, 
around which the dispersive bands exhibit linear dispersion.
The analytic derivation of this critical value is shown in the Supplemental Material~\cite{SM}.
For $t_2/t_1 >  2$ [Fig.~\ref{fig:band}(e)], 
the system becomes an unconventional gapless semiconductor, where the top of the valence band touches the empty flat band.
The Fermi level is right at the flat band, meaning that the DOS is divergingly large.
This will be a source of exotic physical properties, as we will discuss later.
Notably, the extended martini model provides a natural realization of an unconventional gapless semiconductor at half-filling, which is distinct from other typical flat-band lattices.
To be specific, the flat band is at the top or bottom of the entire bands for a line-graph lattice,
while the flat band is half-filled in the Lieb-type lattices. 
Hence, neither of them realizes an unconventional gapless semiconductor.
Considering that the blue bonds are shorter than the red bonds in Fig.~\ref{geometry}(a), 
it is reasonable to assume that $|t_2| \gg |t_1| = |t_1^\prime|$ holds when implementing this structure by $\pi$-electron networks.
We thus focus on the unconventional gapless state [Fig.~\ref{fig:band}(e)] in the following discussions. 

\textit{Two types of unconventional gapless semiconductor.} 
Before proceeding to the material realizations,
we note that the sign of $t_1$ is essential in determining the actual electronic state~\cite{remark}.
The schematic figure of the electronic structures around the Fermi level for $|t_2| > 2 |t_1|$ is shown in Fig.~\ref{fig:types}.
For $t_1 < 0$, which is the case of Fig.~\ref{fig:band}(e),
the dispersive band touching the flat band is convex downward 
and is completely filled, whereas the flat band is completely empty.
We refer to this case as type-I [Fig.~\ref{fig:types}(a)]. 
Meanwhile, for $t_1 > 0$, 
the entire band structure is obtained by flipping the sign of the eigenenergies of those 
for $t_1 < 0$. 
As a result, we find that the dispersive band touching the flat band is convex upward and is completely empty;
instead, the flat band right below the dispersive band is completely filled. 
We refer to this case as type-II [Fig.~\ref{fig:types}(b)].
In fact, both types are feasible by the choice of mother materials and adatoms, as we shall discuss below. 

\textit{Material design of an extended martini lattice.} 
We now argue the materials realization of the extended martini model.
Geometric and electronic structures of realistic martini structures derived from 
graphene and silicene are investigated using the density functional theory~\cite{dft1,dft2}. 
See Supplemental Material~\cite{SM} (and references~\onlinecite{state1,state3,gga,ppot} therein) for details of the calculation methods.

Covalent honeycomb networks of C and Si are plausible 
starting materials to design an extended martini lattice
because a corresponding network is obtained by partially 
thinning out the $\pi$ electrons by adsorption  
or chemical substitution.
Here, we focus on the adsorption on graphene, leaving the results for silicene for the Supplemental Material~\cite{SM}.
Figure~\ref{geometry}(a) shows the schematic views of 
the possible structure of an extended martini lattice derived from a honeycomb C network.
An extended martini lattice can be found in Fig.~\ref{geometry}(a) 
as downward three-pointed stars comprising white circles by adsorbing or 
substituting four of eight atomic sites forming upward three-pointed-star in each 2$\times$2 lateral unit cell by foreign atoms.
The chemisorption of H onto graphene and silicene effectively removes $\pi$ electrons on H terminated C atomic sites, 
leading to an extended martini lattice of $\pi$ electrons on the partially H adsorbed graphene 
[C-H in Fig.~\ref{geometry}(b)].
Partial fluorination of graphene also effectively causes an extended martini lattice [C-F in Fig.~\ref{geometry}(c)].
The optimized lattice parameters of extended martini networks of C-H and C-F are 5.04 and 5.04 {\AA}, 
respectively, which are slightly longer than that of graphene, 
because the chemisorption on four of eight atomic sites per cell leads to sp$^3$ bonds.
In addition to the adsorption, the substitution of C by B and N atoms 
can substantially modulate the $\pi$ electron environment on the honeycomb networks, 
so that the in-plane heterostructures of three-pointed stars of C and B/N are possible candidates for the extended martini lattice.
We argue the details of this case in Supplemental Material~\cite{SM}.
\begin{figure}
\begin{center}
\includegraphics[width=0.93\linewidth]{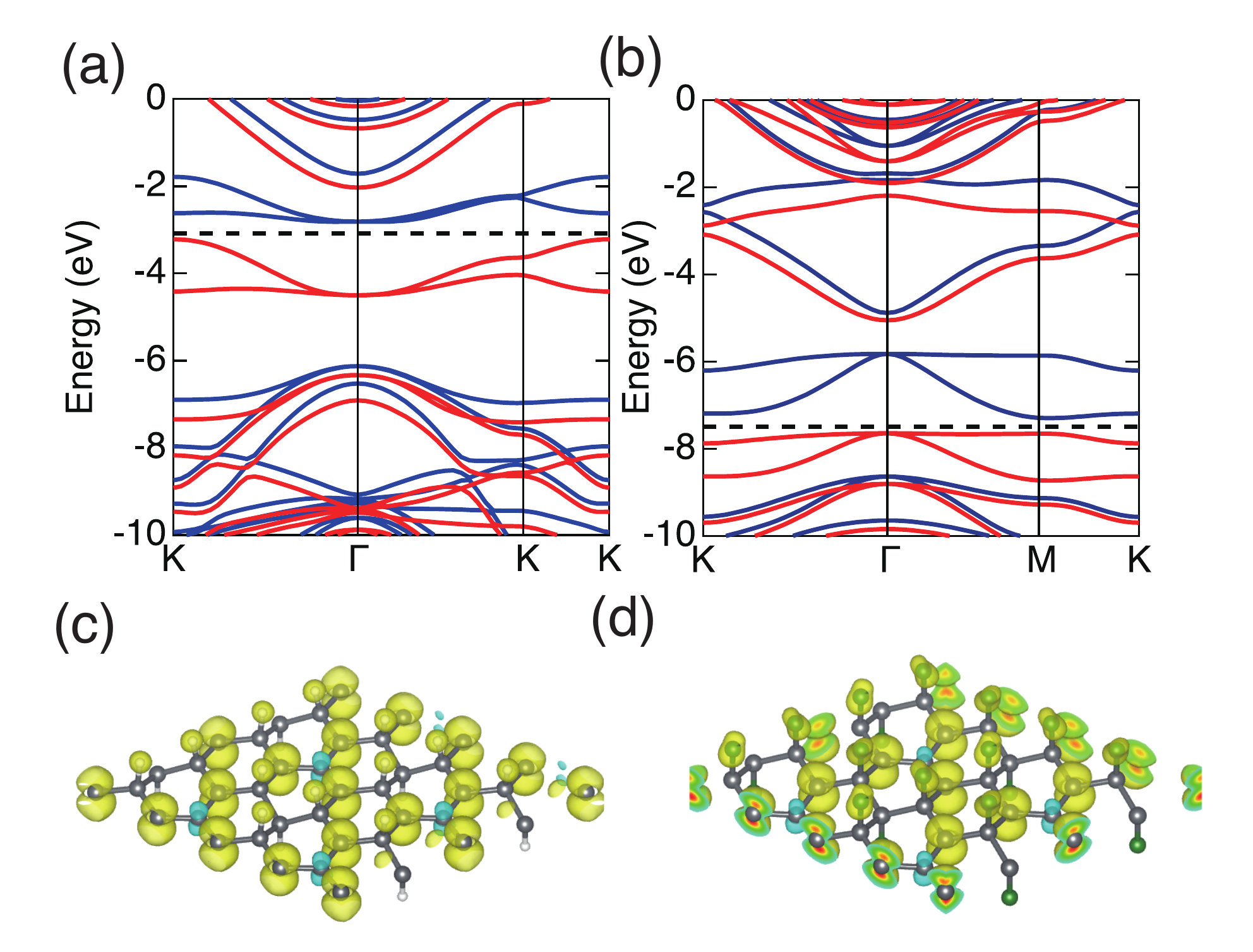}
\caption{
Upper row: Electronic structures of (a) partially hydrogenated graphene, 
and (b) partially fluorinated graphene. 
The energies are measured from the vacuum level.
Red and blue curves denote the energy band for majority and minority states, respectively.
The horizontal dotted line indicates the Fermi level.
Lower row: Isosurfaces of polarized electron spin, $\Delta \rho(\Vec{r})= \rho(\Vec{r})_{mj} - \rho(\Vec{r})_{mn}$, 
where $\rho(\Vec{r})_{mj}$ and $\rho(\Vec{r})_{mn}$ are charge densities of majority and minority spin states, respectively, of (c) partially hydrogenated graphene, 
and (d) partially fluorinated graphene. 
Yellow and blue isosurfaces denote the electron accumulation and depression, 
respectively. Gray, green, and white spheres denote C, F, and H atoms, respectively.
}
\label{banddft}
\end{center}
\vspace{-10pt}
\end{figure}

Figures~\ref{banddft}(a) and \ref{banddft}(b) 
show the electronic band structure of an extended martini lattice 
derived from the honeycomb networks of C. 
The extended martini systems of C-H and C-F are semiconductors where the two branches 
just below and above the Fermi level in the majority and the minority spin states, 
respectively, possess a characteristic dispersion relation: 
One of the two is less dispersive and the other has dispersion with substantial width. 
Furthermore, these two branches degenerate at the $\Gamma$ point. 
These characteristic dispersion relations are the same as those obtained by the tight-binding approximation, 
indicating that these networks are an extended martini lattice derived from graphene by atom adsorption. 
The adsorbate species on graphene can control the band structure attributed to the extended martini lattice. 
The flat band emerges in the lower branches and upper 
branches of two branches for extended martini lattices of C-F and C-H, respectively. 
Comparing these with the tight-binding model, we find that C-F corresponds to the type-I in Fig.~\ref{fig:types}(a), 
while C-H corresponds to the type-II in Fig.~\ref{fig:types}(b).
These facts imply that the constituent elements of honeycomb networks and the adsorbates can control 
a sign of the effective electron transfer between the next-nearest $\pi$ electrons (i.e., $t_1$ and $t_1^\prime$), 
allowing further band edge engineering in these graphene derivatives. 

In an extended martini lattice derived from graphene, a long-range wave function overlap and incomplete 
$\pi$ termination by adsorbates lead to a small but finite band dispersion in the flat band state.
The calculated band widths of the flat band states are 0.46 and 0.23 eV for 
extended martini lattices of  C-H and C-F, 
respectively.   
This small but finite band width leads to partial occupation that causes a large Fermi level instability. 
Extended martini networks of C-H and C-F exhibit spin polarization as shown in Figs.~\ref{banddft}(a) and \ref{banddft}(b), respectively.
The majority and minority bands associated 
with the martini flat band state shift downward and upward, respectively, owing to the spin polarization. 
The polarized spin is primarily distributed on the three edges of a three-pointed star of bare C atoms 
[Fig.~\ref{banddft}(c) for C-H and Fig.~\ref{banddft}(d) for C-F].
The distribution corresponds to the wave function of the martini bands at $\Gamma$ and the Fermi level.
Namely, 
the flat band wave function of Eq.~(\ref{eq:fbwf}) has a vanishing amplitude at sublattice 1.
The number of polarized electrons is 2 per 2$\times$2 unit cell for both C-H and C-F,
corresponding to 0.15 $\mu_B$/{\AA}$^2$. 
Therefore, these facts indicate that the partial hydrogenation or fluorination of graphene,
(as well as the partial hydrogenation of silicene~\cite{SM}),
are magnetic materials 
whose magnetization is attributed to itinerant 
$\pi$ electrons on atomic layer materials. 
In contrast, C/BN heterostructures obtained by 
B/N substitution do not exhibit spin polarization, 
owing to the substantial band width of the flat band states. 
The calculated band widths of the flat band states are 1 eV or wider~\cite{SM}.

\textit{Summary and discussions.} 
We have investigated the characteristic electronic structures in an extended martini lattice model and its 
materialization in partially chemisorbed graphene and silicene.
The analytic treatment of the tight-binding model reveals that an unconventional gapless semiconductor with 
the exact flat band at the Fermi level is realized when $|t_2| > 2 |t_1|$ (for $t_1 = t_1^\prime$).
Depending on the sign of $t_1$ and $t_1^\prime$, 
the gapless semiconductor is classified into a type-I, with the completely empty flat band, 
and a type-II, with the completely filled flat band.
In actual materials, the types can be tuned by the species of the adsorbed atoms. 

We close this Letter by addressing possible future 
problems and intriguing functionalities of extended martini materials.
First, regarding the candidate materials design, 
there can be rich combinations of mother compounds, adatoms, and substituents.
The choices of them determine the tight-binding parameters,
as well as the characters out of our idealized models 
such as the band width of nearly flat bands, which may enable the realization of various band structures in Fig.~\ref{fig:band}. 
In addition to monolayer materials, our scheme of electronic state engineering 
is also applicable to surfaces systems, such as the (111) surface of silicon or diamond~\cite{Okada2003,Zhou2017}.
Extensive materials search and electronic-structure analysis by the density functional theory calculation
will be an interesting future problem.
Experimentally, recent developments on the scanning tunneling microscope technique have enabled 
single-atom manipulation~\cite{Lyo1991,Sugimoto2007,Sugimoto2008},
which will open the door to fabricating extended martini materials by a periodic alignment of the adatoms.
Another class of candidate materials is metal organic frameworks and covalent organic frameworks.
There, a three-dimensional analog of the extended martini network, which corresponds to a pyrochlore lattice 
with one additional site per downward tetrahedron may also be pursued.

Second, as for the functionality due to spin polarization, 
it is expected that an electronic state such as that shown in 
Figs.~\ref{banddft}(a) and \ref{banddft}(b) can be utilized for spin-filtered transport.
Furthermore, when the spin-polarization is weak such that the flat band of the majority spin is partially filled,
the interplay between the flat-band state of majority spins and the mobile holes of the minority spins will give rise to exotic many-body states.
Studying such situations will be an intriguing future problem.  

Finally, if one can suppress the spin polarization and 
retain a nonmagnetic unconventional gapless semiconductor, 
the sharply-varying DOS around the Fermi level can be a source of large thermoelectric responses~\cite{Sommerfeld1933,Mott1936,Luttinger1964}. 
Further, very recent theoretical studies have revealed that the 
quadratic band touching between the flat and the dispersive bands
gives rise to an unconventional quantum geometric tensor, 
which anomalously affects various fundamental quantities such as
a magnetic-field response~\cite{Rhim2020,Hwang2021_LL} and a superfluid weight~\cite{Iskin2018,Hu2019,Huhtinen2022}.
The extended martini materials will serve as suitable platforms for studying these phenomena. 
 
\begin{acknowledgements}
The authors thank 
the Japan Science and Technology Agency, Core Research for Evolutionary Science and Technology 
(JST-CREST; Grant No.~JPMJCR1715, No.~JPMJCR19T1, and No.~JPMJCR20B5) and the Japan Society for the Promotion of Science,
Grants-in-Aid for Scientific Research 
(JSPS KAKENHI; Grant No.~JP21H05233, No.~JP21H05232, No.~JP21K14484, No.~JP20K22323, No.~JP20H00316, No.~JP20H02080, No.~JP20K05253, No.~JP20K14371, and No.~JP20H05664), 
and the Joint Research Program on Zero-Emission Energy Research, Institute of Advanced Energy, Kyoto University.
A part of the calculations was performed on an NEC SX-Aurora TSUBASA at the Cybermedia Center at Osaka University. 
\end{acknowledgements}

\bibliographystyle{apsrev4-2}
\bibliography{exmartini}

\begin{thebibliography}{71}%
\makeatletter
\providecommand \@ifxundefined [1]{%
 \@ifx{#1\undefined}
}%
\providecommand \@ifnum [1]{%
 \ifnum #1\expandafter \@firstoftwo
 \else \expandafter \@secondoftwo
 \fi
}%
\providecommand \@ifx [1]{%
 \ifx #1\expandafter \@firstoftwo
 \else \expandafter \@secondoftwo
 \fi
}%
\providecommand \natexlab [1]{#1}%
\providecommand \enquote  [1]{``#1''}%
\providecommand \bibnamefont  [1]{#1}%
\providecommand \bibfnamefont [1]{#1}%
\providecommand \citenamefont [1]{#1}%
\providecommand \href@noop [0]{\@secondoftwo}%
\providecommand \href [0]{\begingroup \@sanitize@url \@href}%
\providecommand \@href[1]{\@@startlink{#1}\@@href}%
\providecommand \@@href[1]{\endgroup#1\@@endlink}%
\providecommand \@sanitize@url [0]{\catcode `\\12\catcode `\$12\catcode
  `\&12\catcode `\#12\catcode `\^12\catcode `\_12\catcode `\%12\relax}%
\providecommand \@@startlink[1]{}%
\providecommand \@@endlink[0]{}%
\providecommand \url  [0]{\begingroup\@sanitize@url \@url }%
\providecommand \@url [1]{\endgroup\@href {#1}{\urlprefix }}%
\providecommand \urlprefix  [0]{URL }%
\providecommand \Eprint [0]{\href }%
\providecommand \doibase [0]{https://doi.org/}%
\providecommand \selectlanguage [0]{\@gobble}%
\providecommand \bibinfo  [0]{\@secondoftwo}%
\providecommand \bibfield  [0]{\@secondoftwo}%
\providecommand \translation [1]{[#1]}%
\providecommand \BibitemOpen [0]{}%
\providecommand \bibitemStop [0]{}%
\providecommand \bibitemNoStop [0]{.\EOS\space}%
\providecommand \EOS [0]{\spacefactor3000\relax}%
\providecommand \BibitemShut  [1]{\csname bibitem#1\endcsname}%
\let\auto@bib@innerbib\@empty
\bibitem [{\citenamefont {Mielke}(1991)}]{Mielke1991}%
  \BibitemOpen
  \bibfield  {author} {\bibinfo {author} {\bibfnamefont {A.}~\bibnamefont
  {Mielke}},\ }\href {https://doi.org/10.1088/0305-4470/24/14/018} {\bibfield
  {journal} {\bibinfo  {journal} {J. Phys. A: Mathematical and General}\
  }\textbf {\bibinfo {volume} {24}},\ \bibinfo {pages} {3311} (\bibinfo {year}
  {1991})}\BibitemShut {NoStop}%
\bibitem [{\citenamefont {Tasaki}(1992)}]{Tasaki1992}%
  \BibitemOpen
  \bibfield  {author} {\bibinfo {author} {\bibfnamefont {H.}~\bibnamefont
  {Tasaki}},\ }\href {https://doi.org/10.1103/PhysRevLett.69.1608} {\bibfield
  {journal} {\bibinfo  {journal} {Phys. Rev. Lett.}\ }\textbf {\bibinfo
  {volume} {69}},\ \bibinfo {pages} {1608} (\bibinfo {year}
  {1992})}\BibitemShut {NoStop}%
\bibitem [{\citenamefont {Imada}\ and\ \citenamefont
  {Kohno}(2000)}]{Imada2000}%
  \BibitemOpen
  \bibfield  {author} {\bibinfo {author} {\bibfnamefont {M.}~\bibnamefont
  {Imada}}\ and\ \bibinfo {author} {\bibfnamefont {M.}~\bibnamefont {Kohno}},\
  }\href {https://doi.org/10.1103/PhysRevLett.84.143} {\bibfield  {journal}
  {\bibinfo  {journal} {Phys. Rev. Lett.}\ }\textbf {\bibinfo {volume} {84}},\
  \bibinfo {pages} {143} (\bibinfo {year} {2000})}\BibitemShut {NoStop}%
\bibitem [{\citenamefont {Heikkil{\"a}}\ \emph {et~al.}(2011)\citenamefont
  {Heikkil{\"a}}, \citenamefont {Kopnin},\ and\ \citenamefont
  {Volovik}}]{Heikkila2011}%
  \BibitemOpen
  \bibfield  {author} {\bibinfo {author} {\bibfnamefont {T.~T.}\ \bibnamefont
  {Heikkil{\"a}}}, \bibinfo {author} {\bibfnamefont {N.~B.}\ \bibnamefont
  {Kopnin}},\ and\ \bibinfo {author} {\bibfnamefont {G.~E.}\ \bibnamefont
  {Volovik}},\ }\href {https://doi.org/10.1134/S0021364011150045} {\bibfield
  {journal} {\bibinfo  {journal} {JETP Letters}\ }\textbf {\bibinfo {volume}
  {94}},\ \bibinfo {pages} {233} (\bibinfo {year} {2011})}\BibitemShut
  {NoStop}%
\bibitem [{\citenamefont {Peotta}\ and\ \citenamefont
  {T{\"o}rm{\"a}}(2015)}]{Peotta2015}%
  \BibitemOpen
  \bibfield  {author} {\bibinfo {author} {\bibfnamefont {S.}~\bibnamefont
  {Peotta}}\ and\ \bibinfo {author} {\bibfnamefont {P.}~\bibnamefont
  {T{\"o}rm{\"a}}},\ }\href {https://doi.org/10.1038/ncomms9944} {\bibfield
  {journal} {\bibinfo  {journal} {Nature Communications}\ }\textbf {\bibinfo
  {volume} {6}},\ \bibinfo {pages} {8944} (\bibinfo {year} {2015})}\BibitemShut
  {NoStop}%
\bibitem [{\citenamefont {Aoki}(2020)}]{Aoki2020}%
  \BibitemOpen
  \bibfield  {author} {\bibinfo {author} {\bibfnamefont {H.}~\bibnamefont
  {Aoki}},\ }\href {https://doi.org/10.1007/s10948-020-05474-6} {\bibfield
  {journal} {\bibinfo  {journal} {Journal of Superconductivity and Novel
  Magnetism}\ }\textbf {\bibinfo {volume} {33}},\ \bibinfo {pages} {2341}
  (\bibinfo {year} {2020})}\BibitemShut {NoStop}%
\bibitem [{\citenamefont {Peri}\ \emph {et~al.}(2021)\citenamefont {Peri},
  \citenamefont {Song}, \citenamefont {Bernevig},\ and\ \citenamefont
  {Huber}}]{Peri2021}%
  \BibitemOpen
  \bibfield  {author} {\bibinfo {author} {\bibfnamefont {V.}~\bibnamefont
  {Peri}}, \bibinfo {author} {\bibfnamefont {Z.-D.}\ \bibnamefont {Song}},
  \bibinfo {author} {\bibfnamefont {B.~A.}\ \bibnamefont {Bernevig}},\ and\
  \bibinfo {author} {\bibfnamefont {S.~D.}\ \bibnamefont {Huber}},\ }\href
  {https://doi.org/10.1103/PhysRevLett.126.027002} {\bibfield  {journal}
  {\bibinfo  {journal} {Phys. Rev. Lett.}\ }\textbf {\bibinfo {volume} {126}},\
  \bibinfo {pages} {027002} (\bibinfo {year} {2021})}\BibitemShut {NoStop}%
\bibitem [{\citenamefont {C{\u{a}}lug{\u{a}}ru}\ \emph
  {et~al.}(2022)\citenamefont {C{\u{a}}lug{\u{a}}ru}, \citenamefont {Chew},
  \citenamefont {Elcoro}, \citenamefont {Xu}, \citenamefont {Regnault},
  \citenamefont {Song},\ and\ \citenamefont {Bernevig}}]{Calugaru2022}%
  \BibitemOpen
  \bibfield  {author} {\bibinfo {author} {\bibfnamefont {D.}~\bibnamefont
  {C{\u{a}}lug{\u{a}}ru}}, \bibinfo {author} {\bibfnamefont {A.}~\bibnamefont
  {Chew}}, \bibinfo {author} {\bibfnamefont {L.}~\bibnamefont {Elcoro}},
  \bibinfo {author} {\bibfnamefont {Y.}~\bibnamefont {Xu}}, \bibinfo {author}
  {\bibfnamefont {N.}~\bibnamefont {Regnault}}, \bibinfo {author}
  {\bibfnamefont {Z.-D.}\ \bibnamefont {Song}},\ and\ \bibinfo {author}
  {\bibfnamefont {B.~A.}\ \bibnamefont {Bernevig}},\ }\href
  {https://doi.org/10.1038/s41567-021-01445-3} {\bibfield  {journal} {\bibinfo
  {journal} {Nature Physics}\ }\textbf {\bibinfo {volume} {18}},\ \bibinfo
  {pages} {185} (\bibinfo {year} {2022})}\BibitemShut {NoStop}%
\bibitem [{\citenamefont {Regnault}\ \emph {et~al.}(2022)\citenamefont
  {Regnault}, \citenamefont {Xu}, \citenamefont {Li}, \citenamefont {Ma},
  \citenamefont {Jovanovic}, \citenamefont {Yazdani}, \citenamefont {Parkin},
  \citenamefont {Felser}, \citenamefont {Schoop}, \citenamefont {Ong},
  \citenamefont {Cava}, \citenamefont {Elcoro}, \citenamefont {Song},\ and\
  \citenamefont {Bernevig}}]{Regnault2022}%
  \BibitemOpen
  \bibfield  {author} {\bibinfo {author} {\bibfnamefont {N.}~\bibnamefont
  {Regnault}}, \bibinfo {author} {\bibfnamefont {Y.}~\bibnamefont {Xu}},
  \bibinfo {author} {\bibfnamefont {M.-R.}\ \bibnamefont {Li}}, \bibinfo
  {author} {\bibfnamefont {D.-S.}\ \bibnamefont {Ma}}, \bibinfo {author}
  {\bibfnamefont {M.}~\bibnamefont {Jovanovic}}, \bibinfo {author}
  {\bibfnamefont {A.}~\bibnamefont {Yazdani}}, \bibinfo {author} {\bibfnamefont
  {S.~S.~P.}\ \bibnamefont {Parkin}}, \bibinfo {author} {\bibfnamefont
  {C.}~\bibnamefont {Felser}}, \bibinfo {author} {\bibfnamefont {L.~M.}\
  \bibnamefont {Schoop}}, \bibinfo {author} {\bibfnamefont {N.~P.}\
  \bibnamefont {Ong}}, \bibinfo {author} {\bibfnamefont {R.~J.}\ \bibnamefont
  {Cava}}, \bibinfo {author} {\bibfnamefont {L.}~\bibnamefont {Elcoro}},
  \bibinfo {author} {\bibfnamefont {Z.-D.}\ \bibnamefont {Song}},\ and\
  \bibinfo {author} {\bibfnamefont {B.~A.}\ \bibnamefont {Bernevig}},\ }\href
  {https://doi.org/10.1038/s41586-022-04519-1} {\bibfield  {journal} {\bibinfo
  {journal} {Nature}\ }\textbf {\bibinfo {volume} {603}},\ \bibinfo {pages}
  {824} (\bibinfo {year} {2022})}\BibitemShut {NoStop}%
\bibitem [{\citenamefont {Chiu}\ \emph {et~al.}(2022)\citenamefont {Chiu},
  \citenamefont {Carroll}, \citenamefont {Regnault},\ and\ \citenamefont
  {Houck}}]{Chiu2022}%
  \BibitemOpen
  \bibfield  {author} {\bibinfo {author} {\bibfnamefont {C.~S.}\ \bibnamefont
  {Chiu}}, \bibinfo {author} {\bibfnamefont {A.~N.}\ \bibnamefont {Carroll}},
  \bibinfo {author} {\bibfnamefont {N.}~\bibnamefont {Regnault}},\ and\
  \bibinfo {author} {\bibfnamefont {A.~A.}\ \bibnamefont {Houck}},\ }\href
  {https://doi.org/10.1103/PhysRevResearch.4.023063} {\bibfield  {journal}
  {\bibinfo  {journal} {Phys. Rev. Research}\ }\textbf {\bibinfo {volume}
  {4}},\ \bibinfo {pages} {023063} (\bibinfo {year} {2022})}\BibitemShut
  {NoStop}%
\bibitem [{\citenamefont {Wallace}(1947)}]{Wallace1947}%
  \BibitemOpen
  \bibfield  {author} {\bibinfo {author} {\bibfnamefont {P.~R.}\ \bibnamefont
  {Wallace}},\ }\href {https://doi.org/10.1103/PhysRev.71.622} {\bibfield
  {journal} {\bibinfo  {journal} {Phys. Rev.}\ }\textbf {\bibinfo {volume}
  {71}},\ \bibinfo {pages} {622} (\bibinfo {year} {1947})}\BibitemShut
  {NoStop}%
\bibitem [{\citenamefont {Novoselov}\ \emph {et~al.}(2004)\citenamefont
  {Novoselov}, \citenamefont {Geim}, \citenamefont {Morozov}, \citenamefont
  {Jiang}, \citenamefont {Zhang}, \citenamefont {Dubonos}, \citenamefont
  {Grigorieva},\ and\ \citenamefont {Firsov}}]{Novoselov2004}%
  \BibitemOpen
  \bibfield  {author} {\bibinfo {author} {\bibfnamefont {K.~S.}\ \bibnamefont
  {Novoselov}}, \bibinfo {author} {\bibfnamefont {A.~K.}\ \bibnamefont {Geim}},
  \bibinfo {author} {\bibfnamefont {S.~V.}\ \bibnamefont {Morozov}}, \bibinfo
  {author} {\bibfnamefont {D.}~\bibnamefont {Jiang}}, \bibinfo {author}
  {\bibfnamefont {Y.}~\bibnamefont {Zhang}}, \bibinfo {author} {\bibfnamefont
  {S.~V.}\ \bibnamefont {Dubonos}}, \bibinfo {author} {\bibfnamefont {I.~V.}\
  \bibnamefont {Grigorieva}},\ and\ \bibinfo {author} {\bibfnamefont {A.~A.}\
  \bibnamefont {Firsov}},\ }\href {https://doi.org/10.1126/science.1102896}
  {\bibfield  {journal} {\bibinfo  {journal} {Science}\ }\textbf {\bibinfo
  {volume} {306}},\ \bibinfo {pages} {666} (\bibinfo {year}
  {2004})}\BibitemShut {NoStop}%
\bibitem [{\citenamefont {Novoselov}\ \emph {et~al.}(2005)\citenamefont
  {Novoselov}, \citenamefont {Geim}, \citenamefont {Morozov}, \citenamefont
  {Jiang}, \citenamefont {Katsnelson}, \citenamefont {Grigorieva},
  \citenamefont {Dubonos},\ and\ \citenamefont {Firsov}}]{Novoselov2005}%
  \BibitemOpen
  \bibfield  {author} {\bibinfo {author} {\bibfnamefont {K.~S.}\ \bibnamefont
  {Novoselov}}, \bibinfo {author} {\bibfnamefont {A.~K.}\ \bibnamefont {Geim}},
  \bibinfo {author} {\bibfnamefont {S.~V.}\ \bibnamefont {Morozov}}, \bibinfo
  {author} {\bibfnamefont {D.}~\bibnamefont {Jiang}}, \bibinfo {author}
  {\bibfnamefont {M.~I.}\ \bibnamefont {Katsnelson}}, \bibinfo {author}
  {\bibfnamefont {I.~V.}\ \bibnamefont {Grigorieva}}, \bibinfo {author}
  {\bibfnamefont {S.~V.}\ \bibnamefont {Dubonos}},\ and\ \bibinfo {author}
  {\bibfnamefont {A.~A.}\ \bibnamefont {Firsov}},\ }\href
  {https://doi.org/10.1038/nature04233} {\bibfield  {journal} {\bibinfo
  {journal} {Nature}\ }\textbf {\bibinfo {volume} {438}},\ \bibinfo {pages}
  {197} (\bibinfo {year} {2005})}\BibitemShut {NoStop}%
\bibitem [{\citenamefont {Geim}\ and\ \citenamefont
  {Novoselov}(2007)}]{Geim2007}%
  \BibitemOpen
  \bibfield  {author} {\bibinfo {author} {\bibfnamefont {A.~K.}\ \bibnamefont
  {Geim}}\ and\ \bibinfo {author} {\bibfnamefont {K.~S.}\ \bibnamefont
  {Novoselov}},\ }\href {https://doi.org/10.1038/nmat1849} {\bibfield
  {journal} {\bibinfo  {journal} {Nature Materials}\ }\textbf {\bibinfo
  {volume} {6}},\ \bibinfo {pages} {183} (\bibinfo {year} {2007})}\BibitemShut
  {NoStop}%
\bibitem [{\citenamefont {Castro~Neto}\ \emph {et~al.}(2009)\citenamefont
  {Castro~Neto}, \citenamefont {Guinea}, \citenamefont {Peres}, \citenamefont
  {Novoselov},\ and\ \citenamefont {Geim}}]{CastroNeto2009}%
  \BibitemOpen
  \bibfield  {author} {\bibinfo {author} {\bibfnamefont {A.~H.}\ \bibnamefont
  {Castro~Neto}}, \bibinfo {author} {\bibfnamefont {F.}~\bibnamefont {Guinea}},
  \bibinfo {author} {\bibfnamefont {N.~M.~R.}\ \bibnamefont {Peres}}, \bibinfo
  {author} {\bibfnamefont {K.~S.}\ \bibnamefont {Novoselov}},\ and\ \bibinfo
  {author} {\bibfnamefont {A.~K.}\ \bibnamefont {Geim}},\ }\href
  {https://doi.org/10.1103/RevModPhys.81.109} {\bibfield  {journal} {\bibinfo
  {journal} {Rev. Mod. Phys.}\ }\textbf {\bibinfo {volume} {81}},\ \bibinfo
  {pages} {109} (\bibinfo {year} {2009})}\BibitemShut {NoStop}%
\bibitem [{\citenamefont {Garcia}\ \emph {et~al.}(2011)\citenamefont {Garcia},
  \citenamefont {de~Lima}, \citenamefont {Assali},\ and\ \citenamefont
  {Justo}}]{Garcia2011}%
  \BibitemOpen
  \bibfield  {author} {\bibinfo {author} {\bibfnamefont {J.~C.}\ \bibnamefont
  {Garcia}}, \bibinfo {author} {\bibfnamefont {D.~B.}\ \bibnamefont {de~Lima}},
  \bibinfo {author} {\bibfnamefont {L.~V.~C.}\ \bibnamefont {Assali}},\ and\
  \bibinfo {author} {\bibfnamefont {J.~F.}\ \bibnamefont {Justo}},\ }\href
  {https://doi.org/10.1021/jp203657w} {\bibfield  {journal} {\bibinfo
  {journal} {The Journal of Physical Chemistry C}\ }\textbf {\bibinfo {volume}
  {115}},\ \bibinfo {pages} {13242} (\bibinfo {year} {2011})}\BibitemShut
  {NoStop}%
\bibitem [{\citenamefont {Vogt}\ \emph {et~al.}(2012)\citenamefont {Vogt},
  \citenamefont {De~Padova}, \citenamefont {Quaresima}, \citenamefont {Avila},
  \citenamefont {Frantzeskakis}, \citenamefont {Asensio}, \citenamefont
  {Resta}, \citenamefont {Ealet},\ and\ \citenamefont {Le~Lay}}]{Vogt2012}%
  \BibitemOpen
  \bibfield  {author} {\bibinfo {author} {\bibfnamefont {P.}~\bibnamefont
  {Vogt}}, \bibinfo {author} {\bibfnamefont {P.}~\bibnamefont {De~Padova}},
  \bibinfo {author} {\bibfnamefont {C.}~\bibnamefont {Quaresima}}, \bibinfo
  {author} {\bibfnamefont {J.}~\bibnamefont {Avila}}, \bibinfo {author}
  {\bibfnamefont {E.}~\bibnamefont {Frantzeskakis}}, \bibinfo {author}
  {\bibfnamefont {M.~C.}\ \bibnamefont {Asensio}}, \bibinfo {author}
  {\bibfnamefont {A.}~\bibnamefont {Resta}}, \bibinfo {author} {\bibfnamefont
  {B.}~\bibnamefont {Ealet}},\ and\ \bibinfo {author} {\bibfnamefont
  {G.}~\bibnamefont {Le~Lay}},\ }\href
  {https://doi.org/10.1103/PhysRevLett.108.155501} {\bibfield  {journal}
  {\bibinfo  {journal} {Phys. Rev. Lett.}\ }\textbf {\bibinfo {volume} {108}},\
  \bibinfo {pages} {155501} (\bibinfo {year} {2012})}\BibitemShut {NoStop}%
\bibitem [{\citenamefont {Fleurence}\ \emph {et~al.}(2012)\citenamefont
  {Fleurence}, \citenamefont {Friedlein}, \citenamefont {Ozaki}, \citenamefont
  {Kawai}, \citenamefont {Wang},\ and\ \citenamefont
  {Yamada-Takamura}}]{Fleurence2012}%
  \BibitemOpen
  \bibfield  {author} {\bibinfo {author} {\bibfnamefont {A.}~\bibnamefont
  {Fleurence}}, \bibinfo {author} {\bibfnamefont {R.}~\bibnamefont
  {Friedlein}}, \bibinfo {author} {\bibfnamefont {T.}~\bibnamefont {Ozaki}},
  \bibinfo {author} {\bibfnamefont {H.}~\bibnamefont {Kawai}}, \bibinfo
  {author} {\bibfnamefont {Y.}~\bibnamefont {Wang}},\ and\ \bibinfo {author}
  {\bibfnamefont {Y.}~\bibnamefont {Yamada-Takamura}},\ }\href
  {https://doi.org/10.1103/PhysRevLett.108.245501} {\bibfield  {journal}
  {\bibinfo  {journal} {Phys. Rev. Lett.}\ }\textbf {\bibinfo {volume} {108}},\
  \bibinfo {pages} {245501} (\bibinfo {year} {2012})}\BibitemShut {NoStop}%
\bibitem [{\citenamefont {Cao}\ \emph {et~al.}(2018{\natexlab{a}})\citenamefont
  {Cao}, \citenamefont {Fatemi}, \citenamefont {Fang}, \citenamefont
  {Watanabe}, \citenamefont {Taniguchi}, \citenamefont {Kaxiras},\ and\
  \citenamefont {Jarillo-Herrero}}]{Cao2018}%
  \BibitemOpen
  \bibfield  {author} {\bibinfo {author} {\bibfnamefont {Y.}~\bibnamefont
  {Cao}}, \bibinfo {author} {\bibfnamefont {V.}~\bibnamefont {Fatemi}},
  \bibinfo {author} {\bibfnamefont {S.}~\bibnamefont {Fang}}, \bibinfo {author}
  {\bibfnamefont {K.}~\bibnamefont {Watanabe}}, \bibinfo {author}
  {\bibfnamefont {T.}~\bibnamefont {Taniguchi}}, \bibinfo {author}
  {\bibfnamefont {E.}~\bibnamefont {Kaxiras}},\ and\ \bibinfo {author}
  {\bibfnamefont {P.}~\bibnamefont {Jarillo-Herrero}},\ }\href
  {https://doi.org/10.1038/nature26160} {\bibfield  {journal} {\bibinfo
  {journal} {Nature}\ }\textbf {\bibinfo {volume} {556}},\ \bibinfo {pages}
  {43} (\bibinfo {year} {2018}{\natexlab{a}})}\BibitemShut {NoStop}%
\bibitem [{\citenamefont {Cao}\ \emph {et~al.}(2018{\natexlab{b}})\citenamefont
  {Cao}, \citenamefont {Fatemi}, \citenamefont {Demir}, \citenamefont {Fang},
  \citenamefont {Tomarken}, \citenamefont {Luo}, \citenamefont
  {Sanchez-Yamagishi}, \citenamefont {Watanabe}, \citenamefont {Taniguchi},
  \citenamefont {Kaxiras}, \citenamefont {Ashoori},\ and\ \citenamefont
  {Jarillo-Herrero}}]{Cao2018_2}%
  \BibitemOpen
  \bibfield  {author} {\bibinfo {author} {\bibfnamefont {Y.}~\bibnamefont
  {Cao}}, \bibinfo {author} {\bibfnamefont {V.}~\bibnamefont {Fatemi}},
  \bibinfo {author} {\bibfnamefont {A.}~\bibnamefont {Demir}}, \bibinfo
  {author} {\bibfnamefont {S.}~\bibnamefont {Fang}}, \bibinfo {author}
  {\bibfnamefont {S.~L.}\ \bibnamefont {Tomarken}}, \bibinfo {author}
  {\bibfnamefont {J.~Y.}\ \bibnamefont {Luo}}, \bibinfo {author} {\bibfnamefont
  {J.~D.}\ \bibnamefont {Sanchez-Yamagishi}}, \bibinfo {author} {\bibfnamefont
  {K.}~\bibnamefont {Watanabe}}, \bibinfo {author} {\bibfnamefont
  {T.}~\bibnamefont {Taniguchi}}, \bibinfo {author} {\bibfnamefont
  {E.}~\bibnamefont {Kaxiras}}, \bibinfo {author} {\bibfnamefont {R.~C.}\
  \bibnamefont {Ashoori}},\ and\ \bibinfo {author} {\bibfnamefont
  {P.}~\bibnamefont {Jarillo-Herrero}},\ }\href
  {https://doi.org/10.1038/nature26154} {\bibfield  {journal} {\bibinfo
  {journal} {Nature}\ }\textbf {\bibinfo {volume} {556}},\ \bibinfo {pages}
  {80} (\bibinfo {year} {2018}{\natexlab{b}})}\BibitemShut {NoStop}%
\bibitem [{\citenamefont {Yankowitz}\ \emph {et~al.}(2019)\citenamefont
  {Yankowitz}, \citenamefont {Chen}, \citenamefont {Polshyn}, \citenamefont
  {Zhang}, \citenamefont {Watanabe}, \citenamefont {Taniguchi}, \citenamefont
  {Graf}, \citenamefont {Young},\ and\ \citenamefont {Dean}}]{Yankowitz2019}%
  \BibitemOpen
  \bibfield  {author} {\bibinfo {author} {\bibfnamefont {M.}~\bibnamefont
  {Yankowitz}}, \bibinfo {author} {\bibfnamefont {S.}~\bibnamefont {Chen}},
  \bibinfo {author} {\bibfnamefont {H.}~\bibnamefont {Polshyn}}, \bibinfo
  {author} {\bibfnamefont {Y.}~\bibnamefont {Zhang}}, \bibinfo {author}
  {\bibfnamefont {K.}~\bibnamefont {Watanabe}}, \bibinfo {author}
  {\bibfnamefont {T.}~\bibnamefont {Taniguchi}}, \bibinfo {author}
  {\bibfnamefont {D.}~\bibnamefont {Graf}}, \bibinfo {author} {\bibfnamefont
  {A.~F.}\ \bibnamefont {Young}},\ and\ \bibinfo {author} {\bibfnamefont
  {C.~R.}\ \bibnamefont {Dean}},\ }\href
  {https://doi.org/10.1126/science.aav1910} {\bibfield  {journal} {\bibinfo
  {journal} {Science}\ }\textbf {\bibinfo {volume} {363}},\ \bibinfo {pages}
  {1059} (\bibinfo {year} {2019})},\ \Eprint
  {https://arxiv.org/abs/https://www.science.org/doi/pdf/10.1126/science.aav1910}
  {https://www.science.org/doi/pdf/10.1126/science.aav1910} \BibitemShut
  {NoStop}%
\bibitem [{\citenamefont {Park}\ \emph {et~al.}(2021)\citenamefont {Park},
  \citenamefont {Cao}, \citenamefont {Watanabe}, \citenamefont {Taniguchi},\
  and\ \citenamefont {Jarillo-Herrero}}]{Park2021}%
  \BibitemOpen
  \bibfield  {author} {\bibinfo {author} {\bibfnamefont {J.~M.}\ \bibnamefont
  {Park}}, \bibinfo {author} {\bibfnamefont {Y.}~\bibnamefont {Cao}}, \bibinfo
  {author} {\bibfnamefont {K.}~\bibnamefont {Watanabe}}, \bibinfo {author}
  {\bibfnamefont {T.}~\bibnamefont {Taniguchi}},\ and\ \bibinfo {author}
  {\bibfnamefont {P.}~\bibnamefont {Jarillo-Herrero}},\ }\href
  {https://doi.org/10.1038/s41586-021-03192-0} {\bibfield  {journal} {\bibinfo
  {journal} {Nature}\ }\textbf {\bibinfo {volume} {590}},\ \bibinfo {pages}
  {249} (\bibinfo {year} {2021})}\BibitemShut {NoStop}%
\bibitem [{\citenamefont {Tarnopolsky}\ \emph {et~al.}(2019)\citenamefont
  {Tarnopolsky}, \citenamefont {Kruchkov},\ and\ \citenamefont
  {Vishwanath}}]{Tarnopolsky2019}%
  \BibitemOpen
  \bibfield  {author} {\bibinfo {author} {\bibfnamefont {G.}~\bibnamefont
  {Tarnopolsky}}, \bibinfo {author} {\bibfnamefont {A.~J.}\ \bibnamefont
  {Kruchkov}},\ and\ \bibinfo {author} {\bibfnamefont {A.}~\bibnamefont
  {Vishwanath}},\ }\href {https://doi.org/10.1103/PhysRevLett.122.106405}
  {\bibfield  {journal} {\bibinfo  {journal} {Phys. Rev. Lett.}\ }\textbf
  {\bibinfo {volume} {122}},\ \bibinfo {pages} {106405} (\bibinfo {year}
  {2019})}\BibitemShut {NoStop}%
\bibitem [{\citenamefont {Shima}\ and\ \citenamefont {Aoki}(1993)}]{Shima1993}%
  \BibitemOpen
  \bibfield  {author} {\bibinfo {author} {\bibfnamefont {N.}~\bibnamefont
  {Shima}}\ and\ \bibinfo {author} {\bibfnamefont {H.}~\bibnamefont {Aoki}},\
  }\href {https://doi.org/10.1103/PhysRevLett.71.4389} {\bibfield  {journal}
  {\bibinfo  {journal} {Phys. Rev. Lett.}\ }\textbf {\bibinfo {volume} {71}},\
  \bibinfo {pages} {4389} (\bibinfo {year} {1993})}\BibitemShut {NoStop}%
\bibitem [{\citenamefont {Maruyama}\ \emph {et~al.}(2016)\citenamefont
  {Maruyama}, \citenamefont {Cuong},\ and\ \citenamefont
  {Okada}}]{Maruyama2016}%
  \BibitemOpen
  \bibfield  {author} {\bibinfo {author} {\bibfnamefont {M.}~\bibnamefont
  {Maruyama}}, \bibinfo {author} {\bibfnamefont {N.~T.}\ \bibnamefont
  {Cuong}},\ and\ \bibinfo {author} {\bibfnamefont {S.}~\bibnamefont {Okada}},\
  }\href {https://doi.org/https://doi.org/10.1016/j.carbon.2016.08.090}
  {\bibfield  {journal} {\bibinfo  {journal} {Carbon}\ }\textbf {\bibinfo
  {volume} {109}},\ \bibinfo {pages} {755 } (\bibinfo {year}
  {2016})}\BibitemShut {NoStop}%
\bibitem [{\citenamefont {Maruyama}\ and\ \citenamefont
  {Okada}(2017)}]{Maruyama2017}%
  \BibitemOpen
  \bibfield  {author} {\bibinfo {author} {\bibfnamefont {M.}~\bibnamefont
  {Maruyama}}\ and\ \bibinfo {author} {\bibfnamefont {S.}~\bibnamefont
  {Okada}},\ }\href
  {https://doi.org/https://doi.org/10.1016/j.carbon.2017.08.040} {\bibfield
  {journal} {\bibinfo  {journal} {Carbon}\ }\textbf {\bibinfo {volume} {125}},\
  \bibinfo {pages} {530 } (\bibinfo {year} {2017})}\BibitemShut {NoStop}%
\bibitem [{\citenamefont {Sorimachi}\ and\ \citenamefont
  {Okada}(2017)}]{Sorimachi2017}%
  \BibitemOpen
  \bibfield  {author} {\bibinfo {author} {\bibfnamefont {J.-y.}\ \bibnamefont
  {Sorimachi}}\ and\ \bibinfo {author} {\bibfnamefont {S.}~\bibnamefont
  {Okada}},\ }\href {https://doi.org/10.1103/PhysRevB.96.024103} {\bibfield
  {journal} {\bibinfo  {journal} {Phys. Rev. B}\ }\textbf {\bibinfo {volume}
  {96}},\ \bibinfo {pages} {024103} (\bibinfo {year} {2017})}\BibitemShut
  {NoStop}%
\bibitem [{\citenamefont {Fujii}\ \emph
  {et~al.}(2018{\natexlab{a}})\citenamefont {Fujii}, \citenamefont {Maruyama},
  \citenamefont {Wakabayashi}, \citenamefont {Nakada},\ and\ \citenamefont
  {Okada}}]{Fujii2018}%
  \BibitemOpen
  \bibfield  {author} {\bibinfo {author} {\bibfnamefont {Y.}~\bibnamefont
  {Fujii}}, \bibinfo {author} {\bibfnamefont {M.}~\bibnamefont {Maruyama}},
  \bibinfo {author} {\bibfnamefont {K.}~\bibnamefont {Wakabayashi}}, \bibinfo
  {author} {\bibfnamefont {K.}~\bibnamefont {Nakada}},\ and\ \bibinfo {author}
  {\bibfnamefont {S.}~\bibnamefont {Okada}},\ }\href
  {https://doi.org/10.7566/JPSJ.87.034704} {\bibfield  {journal} {\bibinfo
  {journal} {Journal of the Physical Society of Japan}\ }\textbf {\bibinfo
  {volume} {87}},\ \bibinfo {pages} {034704} (\bibinfo {year}
  {2018}{\natexlab{a}})}\BibitemShut {NoStop}%
\bibitem [{\citenamefont {Fujii}\ \emph
  {et~al.}(2018{\natexlab{b}})\citenamefont {Fujii}, \citenamefont {Maruyama},\
  and\ \citenamefont {Okada}}]{Fujii2018_2}%
  \BibitemOpen
  \bibfield  {author} {\bibinfo {author} {\bibfnamefont {Y.}~\bibnamefont
  {Fujii}}, \bibinfo {author} {\bibfnamefont {M.}~\bibnamefont {Maruyama}},\
  and\ \bibinfo {author} {\bibfnamefont {S.}~\bibnamefont {Okada}},\ }\href
  {https://doi.org/10.7567/jjap.57.125203} {\bibfield  {journal} {\bibinfo
  {journal} {Japanese Journal of Applied Physics}\ }\textbf {\bibinfo {volume}
  {57}},\ \bibinfo {pages} {125203} (\bibinfo {year}
  {2018}{\natexlab{b}})}\BibitemShut {NoStop}%
\bibitem [{\citenamefont {Miyahara}\ \emph {et~al.}(2005)\citenamefont
  {Miyahara}, \citenamefont {Kubo}, \citenamefont {Ono}, \citenamefont
  {Shimomura},\ and\ \citenamefont {Furukawa}}]{Miyahara2005}%
  \BibitemOpen
  \bibfield  {author} {\bibinfo {author} {\bibfnamefont {S.}~\bibnamefont
  {Miyahara}}, \bibinfo {author} {\bibfnamefont {K.}~\bibnamefont {Kubo}},
  \bibinfo {author} {\bibfnamefont {H.}~\bibnamefont {Ono}}, \bibinfo {author}
  {\bibfnamefont {Y.}~\bibnamefont {Shimomura}},\ and\ \bibinfo {author}
  {\bibfnamefont {N.}~\bibnamefont {Furukawa}},\ }\href
  {https://doi.org/10.1143/JPSJ.74.1918} {\bibfield  {journal} {\bibinfo
  {journal} {Journal of the Physical Society of Japan}\ }\textbf {\bibinfo
  {volume} {74}},\ \bibinfo {pages} {1918} (\bibinfo {year}
  {2005})}\BibitemShut {NoStop}%
\bibitem [{\citenamefont {Kubo}\ \emph {et~al.}(2006)\citenamefont {Kubo},
  \citenamefont {Hotta}, \citenamefont {Miyahara},\ and\ \citenamefont
  {Furukawa}}]{Kubo2006}%
  \BibitemOpen
  \bibfield  {author} {\bibinfo {author} {\bibfnamefont {K.}~\bibnamefont
  {Kubo}}, \bibinfo {author} {\bibfnamefont {C.}~\bibnamefont {Hotta}},
  \bibinfo {author} {\bibfnamefont {S.}~\bibnamefont {Miyahara}},\ and\
  \bibinfo {author} {\bibfnamefont {N.}~\bibnamefont {Furukawa}},\ }\href
  {https://doi.org/https://doi.org/10.1016/j.physb.2006.01.102} {\bibfield
  {journal} {\bibinfo  {journal} {Physica B: Condensed Matter}\ }\textbf
  {\bibinfo {volume} {378-380}},\ \bibinfo {pages} {273} (\bibinfo {year}
  {2006})},\ \bibinfo {note} {proceedings of the International Conference on
  Strongly Correlated Electron Systems}\BibitemShut {NoStop}%
\bibitem [{\citenamefont {Scullard}(2006)}]{Scullard2006}%
  \BibitemOpen
  \bibfield  {author} {\bibinfo {author} {\bibfnamefont {C.~R.}\ \bibnamefont
  {Scullard}},\ }\href {https://doi.org/10.1103/PhysRevE.73.016107} {\bibfield
  {journal} {\bibinfo  {journal} {Phys. Rev. E}\ }\textbf {\bibinfo {volume}
  {73}},\ \bibinfo {pages} {016107} (\bibinfo {year} {2006})}\BibitemShut
  {NoStop}%
\bibitem [{\citenamefont {McClarty}\ \emph {et~al.}(2020)\citenamefont
  {McClarty}, \citenamefont {Haque}, \citenamefont {Sen},\ and\ \citenamefont
  {Richter}}]{McClarty2020}%
  \BibitemOpen
  \bibfield  {author} {\bibinfo {author} {\bibfnamefont {P.~A.}\ \bibnamefont
  {McClarty}}, \bibinfo {author} {\bibfnamefont {M.}~\bibnamefont {Haque}},
  \bibinfo {author} {\bibfnamefont {A.}~\bibnamefont {Sen}},\ and\ \bibinfo
  {author} {\bibfnamefont {J.}~\bibnamefont {Richter}},\ }\href
  {https://doi.org/10.1103/PhysRevB.102.224303} {\bibfield  {journal} {\bibinfo
   {journal} {Phys. Rev. B}\ }\textbf {\bibinfo {volume} {102}},\ \bibinfo
  {pages} {224303} (\bibinfo {year} {2020})}\BibitemShut {NoStop}%
\bibitem [{\citenamefont {Matsumoto}\ \emph {et~al.}(2023)\citenamefont
  {Matsumoto}, \citenamefont {Mizoguchi},\ and\ \citenamefont
  {Hatsugai}}]{Matsumoto2022}%
  \BibitemOpen
  \bibfield  {author} {\bibinfo {author} {\bibfnamefont {D.}~\bibnamefont
  {Matsumoto}}, \bibinfo {author} {\bibfnamefont {T.}~\bibnamefont
  {Mizoguchi}},\ and\ \bibinfo {author} {\bibfnamefont {Y.}~\bibnamefont
  {Hatsugai}},\ }\href {https://doi.org/10.7566/JPSJ.92.034705} {\bibfield
  {journal} {\bibinfo  {journal} {Journal of the Physical Society of Japan}\
  }\textbf {\bibinfo {volume} {92}},\ \bibinfo {pages} {034705} (\bibinfo
  {year} {2023})}\BibitemShut {NoStop}%
\bibitem [{SM()}]{SM}%
  \BibitemOpen
  \href@noop {} {\ }\bibinfo {note} {See Supplemental Material for the detailed
  numerical and analytical calculations on the tight-binding model, the details
  of the DFT calculation, and the additional DFT results on partially
  hydrogenated silicene and BCN heterostructures, which includes
  Refs.~\onlinecite{state1,state3,gga,ppot}}\BibitemShut {NoStop}%
\bibitem [{\citenamefont {Hatsugai}\ and\ \citenamefont
  {Maruyama}(2011)}]{Hatsugai2011}%
  \BibitemOpen
  \bibfield  {author} {\bibinfo {author} {\bibfnamefont {Y.}~\bibnamefont
  {Hatsugai}}\ and\ \bibinfo {author} {\bibfnamefont {I.}~\bibnamefont
  {Maruyama}},\ }\href {https://doi.org/10.1209/0295-5075/95/20003} {\bibfield
  {journal} {\bibinfo  {journal} {{EPL} (Europhysics Letters)}\ }\textbf
  {\bibinfo {volume} {95}},\ \bibinfo {pages} {20003} (\bibinfo {year}
  {2011})}\BibitemShut {NoStop}%
\bibitem [{\citenamefont {Hatsugai}\ \emph {et~al.}(2015)\citenamefont
  {Hatsugai}, \citenamefont {Shiraishi},\ and\ \citenamefont
  {Aoki}}]{Hatsugai2015}%
  \BibitemOpen
  \bibfield  {author} {\bibinfo {author} {\bibfnamefont {Y.}~\bibnamefont
  {Hatsugai}}, \bibinfo {author} {\bibfnamefont {K.}~\bibnamefont
  {Shiraishi}},\ and\ \bibinfo {author} {\bibfnamefont {H.}~\bibnamefont
  {Aoki}},\ }\href {https://doi.org/10.1088/1367-2630/17/2/025009} {\bibfield
  {journal} {\bibinfo  {journal} {New Journal of Physics}\ }\textbf {\bibinfo
  {volume} {17}},\ \bibinfo {pages} {025009} (\bibinfo {year}
  {2015})}\BibitemShut {NoStop}%
\bibitem [{\citenamefont {Mizoguchi}\ and\ \citenamefont
  {Hatsugai}(2019)}]{Mizoguchi2019}%
  \BibitemOpen
  \bibfield  {author} {\bibinfo {author} {\bibfnamefont {T.}~\bibnamefont
  {Mizoguchi}}\ and\ \bibinfo {author} {\bibfnamefont {Y.}~\bibnamefont
  {Hatsugai}},\ }\href {https://doi.org/10.1209/0295-5075/127/47001} {\bibfield
   {journal} {\bibinfo  {journal} {{EPL} (Europhysics Letters)}\ }\textbf
  {\bibinfo {volume} {127}},\ \bibinfo {pages} {47001} (\bibinfo {year}
  {2019})}\BibitemShut {NoStop}%
\bibitem [{\citenamefont {Mizoguchi}\ \emph {et~al.}(2019)\citenamefont
  {Mizoguchi}, \citenamefont {Maruyama}, \citenamefont {Okada},\ and\
  \citenamefont {Hatsugai}}]{Mizoguchi2019_star}%
  \BibitemOpen
  \bibfield  {author} {\bibinfo {author} {\bibfnamefont {T.}~\bibnamefont
  {Mizoguchi}}, \bibinfo {author} {\bibfnamefont {M.}~\bibnamefont {Maruyama}},
  \bibinfo {author} {\bibfnamefont {S.}~\bibnamefont {Okada}},\ and\ \bibinfo
  {author} {\bibfnamefont {Y.}~\bibnamefont {Hatsugai}},\ }\href
  {https://doi.org/10.1103/PhysRevMaterials.3.114201} {\bibfield  {journal}
  {\bibinfo  {journal} {Phys. Rev. Materials}\ }\textbf {\bibinfo {volume}
  {3}},\ \bibinfo {pages} {114201} (\bibinfo {year} {2019})}\BibitemShut
  {NoStop}%
\bibitem [{\citenamefont {Mizoguchi}\ and\ \citenamefont
  {Hatsugai}(2020)}]{Mizoguchi2020}%
  \BibitemOpen
  \bibfield  {author} {\bibinfo {author} {\bibfnamefont {T.}~\bibnamefont
  {Mizoguchi}}\ and\ \bibinfo {author} {\bibfnamefont {Y.}~\bibnamefont
  {Hatsugai}},\ }\href {https://doi.org/10.1103/PhysRevB.101.235125} {\bibfield
   {journal} {\bibinfo  {journal} {Phys. Rev. B}\ }\textbf {\bibinfo {volume}
  {101}},\ \bibinfo {pages} {235125} (\bibinfo {year} {2020})}\BibitemShut
  {NoStop}%
\bibitem [{\citenamefont {Mizoguchi}\ \emph {et~al.}(2021)\citenamefont
  {Mizoguchi}, \citenamefont {Kuno},\ and\ \citenamefont
  {Hatsugai}}]{Mizoguchi2021_skagome}%
  \BibitemOpen
  \bibfield  {author} {\bibinfo {author} {\bibfnamefont {T.}~\bibnamefont
  {Mizoguchi}}, \bibinfo {author} {\bibfnamefont {Y.}~\bibnamefont {Kuno}},\
  and\ \bibinfo {author} {\bibfnamefont {Y.}~\bibnamefont {Hatsugai}},\ }\href
  {https://doi.org/10.1103/PhysRevB.104.035161} {\bibfield  {journal} {\bibinfo
   {journal} {Phys. Rev. B}\ }\textbf {\bibinfo {volume} {104}},\ \bibinfo
  {pages} {035161} (\bibinfo {year} {2021})}\BibitemShut {NoStop}%
\bibitem [{Rem()}]{Remark1}%
  \BibitemOpen
  \href@noop {} {\ }\bibinfo {note} {Note that $\bm{\psi}_{\bm{k},1}$ and
  $\bm{\psi}_{\bm{k},3}$ are actually $\bm{k}$ independent}\BibitemShut
  {NoStop}%
\bibitem [{\citenamefont {Bergman}\ \emph {et~al.}(2008)\citenamefont
  {Bergman}, \citenamefont {Wu},\ and\ \citenamefont {Balents}}]{Bergman2008}%
  \BibitemOpen
  \bibfield  {author} {\bibinfo {author} {\bibfnamefont {D.~L.}\ \bibnamefont
  {Bergman}}, \bibinfo {author} {\bibfnamefont {C.}~\bibnamefont {Wu}},\ and\
  \bibinfo {author} {\bibfnamefont {L.}~\bibnamefont {Balents}},\ }\href
  {https://doi.org/10.1103/PhysRevB.78.125104} {\bibfield  {journal} {\bibinfo
  {journal} {Phys. Rev. B}\ }\textbf {\bibinfo {volume} {78}},\ \bibinfo
  {pages} {125104} (\bibinfo {year} {2008})}\BibitemShut {NoStop}%
\bibitem [{\citenamefont {Bilitewski}\ and\ \citenamefont
  {Moessner}(2018)}]{Bilitewski2018}%
  \BibitemOpen
  \bibfield  {author} {\bibinfo {author} {\bibfnamefont {T.}~\bibnamefont
  {Bilitewski}}\ and\ \bibinfo {author} {\bibfnamefont {R.}~\bibnamefont
  {Moessner}},\ }\href {https://doi.org/10.1103/PhysRevB.98.235109} {\bibfield
  {journal} {\bibinfo  {journal} {Phys. Rev. B}\ }\textbf {\bibinfo {volume}
  {98}},\ \bibinfo {pages} {235109} (\bibinfo {year} {2018})}\BibitemShut
  {NoStop}%
\bibitem [{\citenamefont {Rhim}\ and\ \citenamefont {Yang}(2019)}]{Rhim2019}%
  \BibitemOpen
  \bibfield  {author} {\bibinfo {author} {\bibfnamefont {J.-W.}\ \bibnamefont
  {Rhim}}\ and\ \bibinfo {author} {\bibfnamefont {B.-J.}\ \bibnamefont
  {Yang}},\ }\href {https://doi.org/10.1103/PhysRevB.99.045107} {\bibfield
  {journal} {\bibinfo  {journal} {Phys. Rev. B}\ }\textbf {\bibinfo {volume}
  {99}},\ \bibinfo {pages} {045107} (\bibinfo {year} {2019})}\BibitemShut
  {NoStop}%
\bibitem [{\citenamefont {Hwang}\ \emph
  {et~al.}(2021{\natexlab{a}})\citenamefont {Hwang}, \citenamefont {Rhim},\
  and\ \citenamefont {Yang}}]{Hwang2021}%
  \BibitemOpen
  \bibfield  {author} {\bibinfo {author} {\bibfnamefont {Y.}~\bibnamefont
  {Hwang}}, \bibinfo {author} {\bibfnamefont {J.-W.}\ \bibnamefont {Rhim}},\
  and\ \bibinfo {author} {\bibfnamefont {B.-J.}\ \bibnamefont {Yang}},\ }\href
  {https://doi.org/10.1103/PhysRevB.104.L081104} {\bibfield  {journal}
  {\bibinfo  {journal} {Phys. Rev. B}\ }\textbf {\bibinfo {volume} {104}},\
  \bibinfo {pages} {L081104} (\bibinfo {year}
  {2021}{\natexlab{a}})}\BibitemShut {NoStop}%
\bibitem [{\citenamefont {Hwang}\ \emph
  {et~al.}(2021{\natexlab{b}})\citenamefont {Hwang}, \citenamefont {Rhim},\
  and\ \citenamefont {Yang}}]{Hwang2021_2}%
  \BibitemOpen
  \bibfield  {author} {\bibinfo {author} {\bibfnamefont {Y.}~\bibnamefont
  {Hwang}}, \bibinfo {author} {\bibfnamefont {J.-W.}\ \bibnamefont {Rhim}},\
  and\ \bibinfo {author} {\bibfnamefont {B.-J.}\ \bibnamefont {Yang}},\ }\href
  {https://doi.org/10.1103/PhysRevB.104.085144} {\bibfield  {journal} {\bibinfo
   {journal} {Phys. Rev. B}\ }\textbf {\bibinfo {volume} {104}},\ \bibinfo
  {pages} {085144} (\bibinfo {year} {2021}{\natexlab{b}})}\BibitemShut
  {NoStop}%
\bibitem [{\citenamefont {Graf}\ and\ \citenamefont
  {Pi\'echon}(2021)}]{Graf2021}%
  \BibitemOpen
  \bibfield  {author} {\bibinfo {author} {\bibfnamefont {A.}~\bibnamefont
  {Graf}}\ and\ \bibinfo {author} {\bibfnamefont {F.}~\bibnamefont
  {Pi\'echon}},\ }\href {https://doi.org/10.1103/PhysRevB.104.195128}
  {\bibfield  {journal} {\bibinfo  {journal} {Phys. Rev. B}\ }\textbf {\bibinfo
  {volume} {104}},\ \bibinfo {pages} {195128} (\bibinfo {year}
  {2021})}\BibitemShut {NoStop}%
\bibitem [{\citenamefont {Hatsugai}(2021)}]{Hatsugai2021}%
  \BibitemOpen
  \bibfield  {author} {\bibinfo {author} {\bibfnamefont {Y.}~\bibnamefont
  {Hatsugai}},\ }\href
  {https://doi.org/https://doi.org/10.1016/j.aop.2021.168453} {\bibfield
  {journal} {\bibinfo  {journal} {Annals of Physics}\ }\textbf {\bibinfo
  {volume} {435}},\ \bibinfo {pages} {168453} (\bibinfo {year} {2021})},\
  \bibinfo {note} {special Issue on Localisation 2020}\BibitemShut {NoStop}%
\bibitem [{\citenamefont {Mizoguchi}\ \emph {et~al.}(2022)\citenamefont
  {Mizoguchi}, \citenamefont {Kuno},\ and\ \citenamefont
  {Hatsugai}}]{Mizoguchi2022}%
  \BibitemOpen
  \bibfield  {author} {\bibinfo {author} {\bibfnamefont {T.}~\bibnamefont
  {Mizoguchi}}, \bibinfo {author} {\bibfnamefont {Y.}~\bibnamefont {Kuno}},\
  and\ \bibinfo {author} {\bibfnamefont {Y.}~\bibnamefont {Hatsugai}},\
  }\bibfield  {journal} {\bibinfo  {journal} {Progress of Theoretical and
  Experimental Physics}\ }\textbf {\bibinfo {volume} {2022}},\ \href
  {https://doi.org/10.1093/ptep/ptac015} {10.1093/ptep/ptac015} (\bibinfo
  {year} {2022}),\ \bibinfo {note} {023I02}\BibitemShut {NoStop}%
\bibitem [{\citenamefont {Kuroda}\ \emph {et~al.}(2022)\citenamefont {Kuroda},
  \citenamefont {Mizoguchi}, \citenamefont {Araki},\ and\ \citenamefont
  {Hatsugai}}]{Kuroda2022}%
  \BibitemOpen
  \bibfield  {author} {\bibinfo {author} {\bibfnamefont {T.}~\bibnamefont
  {Kuroda}}, \bibinfo {author} {\bibfnamefont {T.}~\bibnamefont {Mizoguchi}},
  \bibinfo {author} {\bibfnamefont {H.}~\bibnamefont {Araki}},\ and\ \bibinfo
  {author} {\bibfnamefont {Y.}~\bibnamefont {Hatsugai}},\ }\href
  {https://doi.org/10.7566/JPSJ.91.044703} {\bibfield  {journal} {\bibinfo
  {journal} {Journal of the Physical Society of Japan}\ }\textbf {\bibinfo
  {volume} {91}},\ \bibinfo {pages} {044703} (\bibinfo {year}
  {2022})}\BibitemShut {NoStop}%
\bibitem [{rem()}]{remark}%
  \BibitemOpen
  \href@noop {} {}\bibinfo {note} {It is worth noting that the sign of $t_2$ is
  irrelevant to the band structure. We argue this point in Supplemental
  Material~\cite{SM}.}\BibitemShut {Stop}%
\bibitem [{\citenamefont {Hohenberg}\ and\ \citenamefont {Kohn}(1964)}]{dft1}%
  \BibitemOpen
  \bibfield  {author} {\bibinfo {author} {\bibfnamefont {P.}~\bibnamefont
  {Hohenberg}}\ and\ \bibinfo {author} {\bibfnamefont {W.}~\bibnamefont
  {Kohn}},\ }\href {https://doi.org/10.1103/PhysRev.136.B864} {\bibfield
  {journal} {\bibinfo  {journal} {Phys. Rev.}\ }\textbf {\bibinfo {volume}
  {136}},\ \bibinfo {pages} {B864} (\bibinfo {year} {1964})}\BibitemShut
  {NoStop}%
\bibitem [{\citenamefont {Kohn}\ and\ \citenamefont {Sham}(1965)}]{dft2}%
  \BibitemOpen
  \bibfield  {author} {\bibinfo {author} {\bibfnamefont {W.}~\bibnamefont
  {Kohn}}\ and\ \bibinfo {author} {\bibfnamefont {L.~J.}\ \bibnamefont
  {Sham}},\ }\href {https://doi.org/10.1103/PhysRev.140.A1133} {\bibfield
  {journal} {\bibinfo  {journal} {Phys. Rev.}\ }\textbf {\bibinfo {volume}
  {140}},\ \bibinfo {pages} {A1133} (\bibinfo {year} {1965})}\BibitemShut
  {NoStop}%
\bibitem [{\citenamefont {Y.}\ \emph {et~al.}(2001)\citenamefont {Y.},
  \citenamefont {Iwata},\ and\ \citenamefont {Terakura}}]{state1}%
  \BibitemOpen
  \bibfield  {author} {\bibinfo {author} {\bibnamefont {Y.}}, \bibinfo {author}
  {\bibfnamefont {K.}~\bibnamefont {Iwata}},\ and\ \bibinfo {author}
  {\bibfnamefont {K.}~\bibnamefont {Terakura}},\ }\href
  {https://doi.org/https://doi.org/10.1016/S0169-4332(00)00631-0} {\bibfield
  {journal} {\bibinfo  {journal} {Applied Surface Science}\ }\textbf {\bibinfo
  {volume} {169-170}},\ \bibinfo {pages} {11} (\bibinfo {year}
  {2001})}\BibitemShut {NoStop}%
\bibitem [{sta()}]{state3}%
  \BibitemOpen
  \href@noop {} {}\bibinfo {note} {A simulation tool for atom technology
  (STATE): https://state-doc.readthedocs.io/en/latest/index.html}\BibitemShut
  {NoStop}%
\bibitem [{\citenamefont {Perdew}\ \emph {et~al.}(1996)\citenamefont {Perdew},
  \citenamefont {Burke},\ and\ \citenamefont {Ernzerhof}}]{gga}%
  \BibitemOpen
  \bibfield  {author} {\bibinfo {author} {\bibfnamefont {J.~P.}\ \bibnamefont
  {Perdew}}, \bibinfo {author} {\bibfnamefont {K.}~\bibnamefont {Burke}},\ and\
  \bibinfo {author} {\bibfnamefont {M.}~\bibnamefont {Ernzerhof}},\ }\href
  {https://doi.org/10.1103/PhysRevLett.77.3865} {\bibfield  {journal} {\bibinfo
   {journal} {Phys. Rev. Lett.}\ }\textbf {\bibinfo {volume} {77}},\ \bibinfo
  {pages} {3865} (\bibinfo {year} {1996})}\BibitemShut {NoStop}%
\bibitem [{\citenamefont {Vanderbilt}(1990)}]{ppot}%
  \BibitemOpen
  \bibfield  {author} {\bibinfo {author} {\bibfnamefont {D.}~\bibnamefont
  {Vanderbilt}},\ }\href {https://doi.org/10.1103/PhysRevB.41.7892} {\bibfield
  {journal} {\bibinfo  {journal} {Phys. Rev. B}\ }\textbf {\bibinfo {volume}
  {41}},\ \bibinfo {pages} {7892} (\bibinfo {year} {1990})}\BibitemShut
  {NoStop}%
\bibitem [{\citenamefont {Okada}\ \emph {et~al.}(2003)\citenamefont {Okada},
  \citenamefont {Shiraishi},\ and\ \citenamefont {Oshiyama}}]{Okada2003}%
  \BibitemOpen
  \bibfield  {author} {\bibinfo {author} {\bibfnamefont {S.}~\bibnamefont
  {Okada}}, \bibinfo {author} {\bibfnamefont {K.}~\bibnamefont {Shiraishi}},\
  and\ \bibinfo {author} {\bibfnamefont {A.}~\bibnamefont {Oshiyama}},\ }\href
  {https://doi.org/10.1103/PhysRevLett.90.026803} {\bibfield  {journal}
  {\bibinfo  {journal} {Phys. Rev. Lett.}\ }\textbf {\bibinfo {volume} {90}},\
  \bibinfo {pages} {026803} (\bibinfo {year} {2003})}\BibitemShut {NoStop}%
\bibitem [{\citenamefont {Zhou}\ \emph {et~al.}(2017)\citenamefont {Zhou},
  \citenamefont {Sun},\ and\ \citenamefont {Jena}}]{Zhou2017}%
  \BibitemOpen
  \bibfield  {author} {\bibinfo {author} {\bibfnamefont {J.}~\bibnamefont
  {Zhou}}, \bibinfo {author} {\bibfnamefont {Q.}~\bibnamefont {Sun}},\ and\
  \bibinfo {author} {\bibfnamefont {P.}~\bibnamefont {Jena}},\ }\href
  {https://doi.org/10.1103/PhysRevLett.119.046403} {\bibfield  {journal}
  {\bibinfo  {journal} {Phys. Rev. Lett.}\ }\textbf {\bibinfo {volume} {119}},\
  \bibinfo {pages} {046403} (\bibinfo {year} {2017})}\BibitemShut {NoStop}%
\bibitem [{\citenamefont {Lyo}\ and\ \citenamefont {Avouris}(1991)}]{Lyo1991}%
  \BibitemOpen
  \bibfield  {author} {\bibinfo {author} {\bibfnamefont {I.-W.}\ \bibnamefont
  {Lyo}}\ and\ \bibinfo {author} {\bibfnamefont {P.}~\bibnamefont {Avouris}},\
  }\href {https://doi.org/10.1126/science.253.5016.173} {\bibfield  {journal}
  {\bibinfo  {journal} {Science}\ }\textbf {\bibinfo {volume} {253}},\ \bibinfo
  {pages} {173} (\bibinfo {year} {1991})}\BibitemShut {NoStop}%
\bibitem [{\citenamefont {Sugimoto}\ \emph {et~al.}(2007)\citenamefont
  {Sugimoto}, \citenamefont {Jelinek}, \citenamefont {Pou}, \citenamefont
  {Abe}, \citenamefont {Morita}, \citenamefont {Perez},\ and\ \citenamefont
  {Custance}}]{Sugimoto2007}%
  \BibitemOpen
  \bibfield  {author} {\bibinfo {author} {\bibfnamefont {Y.}~\bibnamefont
  {Sugimoto}}, \bibinfo {author} {\bibfnamefont {P.}~\bibnamefont {Jelinek}},
  \bibinfo {author} {\bibfnamefont {P.}~\bibnamefont {Pou}}, \bibinfo {author}
  {\bibfnamefont {M.}~\bibnamefont {Abe}}, \bibinfo {author} {\bibfnamefont
  {S.}~\bibnamefont {Morita}}, \bibinfo {author} {\bibfnamefont
  {R.}~\bibnamefont {Perez}},\ and\ \bibinfo {author} {\bibfnamefont
  {O.}~\bibnamefont {Custance}},\ }\href
  {https://doi.org/10.1103/PhysRevLett.98.106104} {\bibfield  {journal}
  {\bibinfo  {journal} {Phys. Rev. Lett.}\ }\textbf {\bibinfo {volume} {98}},\
  \bibinfo {pages} {106104} (\bibinfo {year} {2007})}\BibitemShut {NoStop}%
\bibitem [{\citenamefont {Sugimoto}\ \emph {et~al.}(2008)\citenamefont
  {Sugimoto}, \citenamefont {Pou}, \citenamefont {Custance}, \citenamefont
  {Jelinek}, \citenamefont {Abe}, \citenamefont {Perez},\ and\ \citenamefont
  {Morita}}]{Sugimoto2008}%
  \BibitemOpen
  \bibfield  {author} {\bibinfo {author} {\bibfnamefont {Y.}~\bibnamefont
  {Sugimoto}}, \bibinfo {author} {\bibfnamefont {P.}~\bibnamefont {Pou}},
  \bibinfo {author} {\bibfnamefont {O.}~\bibnamefont {Custance}}, \bibinfo
  {author} {\bibfnamefont {P.}~\bibnamefont {Jelinek}}, \bibinfo {author}
  {\bibfnamefont {M.}~\bibnamefont {Abe}}, \bibinfo {author} {\bibfnamefont
  {R.}~\bibnamefont {Perez}},\ and\ \bibinfo {author} {\bibfnamefont
  {S.}~\bibnamefont {Morita}},\ }\href
  {https://doi.org/10.1126/science.1160601} {\bibfield  {journal} {\bibinfo
  {journal} {Science}\ }\textbf {\bibinfo {volume} {322}},\ \bibinfo {pages}
  {413} (\bibinfo {year} {2008})}\BibitemShut {NoStop}%
\bibitem [{\citenamefont {Sommerfeld}\ and\ \citenamefont
  {Bethe}(1933)}]{Sommerfeld1933}%
  \BibitemOpen
  \bibfield  {author} {\bibinfo {author} {\bibfnamefont {A.}~\bibnamefont
  {Sommerfeld}}\ and\ \bibinfo {author} {\bibfnamefont {H.}~\bibnamefont
  {Bethe}},\ }\href@noop {} {\emph {\bibinfo {title} {Elektronentheorie der
  Metalle, Handbuch der Physik}}},\ Vol.\ \bibinfo {volume} {24/2}\ (\bibinfo
  {publisher} {Springer, Berlin/Heidelberg},\ \bibinfo {year}
  {1933})\BibitemShut {NoStop}%
\bibitem [{\citenamefont {Mott}\ and\ \citenamefont {Jones}(1936)}]{Mott1936}%
  \BibitemOpen
  \bibfield  {author} {\bibinfo {author} {\bibfnamefont {N.~F.}\ \bibnamefont
  {Mott}}\ and\ \bibinfo {author} {\bibfnamefont {H.}~\bibnamefont {Jones}},\
  }\href@noop {} {\emph {\bibinfo {title} {The Theory of the Properties of
  Metals and Alloys}}}\ (\bibinfo  {publisher} {Dover New York},\ \bibinfo
  {year} {1936})\BibitemShut {NoStop}%
\bibitem [{\citenamefont {Luttinger}(1964)}]{Luttinger1964}%
  \BibitemOpen
  \bibfield  {author} {\bibinfo {author} {\bibfnamefont {J.~M.}\ \bibnamefont
  {Luttinger}},\ }\href {https://doi.org/10.1103/PhysRev.135.A1505} {\bibfield
  {journal} {\bibinfo  {journal} {Phys. Rev.}\ }\textbf {\bibinfo {volume}
  {135}},\ \bibinfo {pages} {A1505} (\bibinfo {year} {1964})}\BibitemShut
  {NoStop}%
\bibitem [{\citenamefont {Rhim}\ \emph {et~al.}(2020)\citenamefont {Rhim},
  \citenamefont {Kim},\ and\ \citenamefont {Yang}}]{Rhim2020}%
  \BibitemOpen
  \bibfield  {author} {\bibinfo {author} {\bibfnamefont {J.-W.}\ \bibnamefont
  {Rhim}}, \bibinfo {author} {\bibfnamefont {K.}~\bibnamefont {Kim}},\ and\
  \bibinfo {author} {\bibfnamefont {B.-J.}\ \bibnamefont {Yang}},\ }\href
  {https://doi.org/10.1038/s41586-020-2540-1} {\bibfield  {journal} {\bibinfo
  {journal} {Nature}\ }\textbf {\bibinfo {volume} {584}},\ \bibinfo {pages}
  {59} (\bibinfo {year} {2020})}\BibitemShut {NoStop}%
\bibitem [{\citenamefont {Hwang}\ \emph
  {et~al.}(2021{\natexlab{c}})\citenamefont {Hwang}, \citenamefont {Rhim},\
  and\ \citenamefont {Yang}}]{Hwang2021_LL}%
  \BibitemOpen
  \bibfield  {author} {\bibinfo {author} {\bibfnamefont {Y.}~\bibnamefont
  {Hwang}}, \bibinfo {author} {\bibfnamefont {J.-W.}\ \bibnamefont {Rhim}},\
  and\ \bibinfo {author} {\bibfnamefont {B.-J.}\ \bibnamefont {Yang}},\ }\href
  {https://doi.org/10.1038/s41467-021-26765-z} {\bibfield  {journal} {\bibinfo
  {journal} {Nature Communications}\ }\textbf {\bibinfo {volume} {12}},\
  \bibinfo {pages} {6433} (\bibinfo {year} {2021}{\natexlab{c}})}\BibitemShut
  {NoStop}%
\bibitem [{\citenamefont {Iskin}(2018)}]{Iskin2018}%
  \BibitemOpen
  \bibfield  {author} {\bibinfo {author} {\bibfnamefont {M.}~\bibnamefont
  {Iskin}},\ }\href {https://doi.org/10.1103/PhysRevA.97.033625} {\bibfield
  {journal} {\bibinfo  {journal} {Phys. Rev. A}\ }\textbf {\bibinfo {volume}
  {97}},\ \bibinfo {pages} {033625} (\bibinfo {year} {2018})}\BibitemShut
  {NoStop}%
\bibitem [{\citenamefont {Hu}\ \emph {et~al.}(2019)\citenamefont {Hu},
  \citenamefont {Hyart}, \citenamefont {Pikulin},\ and\ \citenamefont
  {Rossi}}]{Hu2019}%
  \BibitemOpen
  \bibfield  {author} {\bibinfo {author} {\bibfnamefont {X.}~\bibnamefont
  {Hu}}, \bibinfo {author} {\bibfnamefont {T.}~\bibnamefont {Hyart}}, \bibinfo
  {author} {\bibfnamefont {D.~I.}\ \bibnamefont {Pikulin}},\ and\ \bibinfo
  {author} {\bibfnamefont {E.}~\bibnamefont {Rossi}},\ }\href
  {https://doi.org/10.1103/PhysRevLett.123.237002} {\bibfield  {journal}
  {\bibinfo  {journal} {Phys. Rev. Lett.}\ }\textbf {\bibinfo {volume} {123}},\
  \bibinfo {pages} {237002} (\bibinfo {year} {2019})}\BibitemShut {NoStop}%
\bibitem [{\citenamefont {Huhtinen}\ \emph {et~al.}(2022)\citenamefont
  {Huhtinen}, \citenamefont {Herzog-Arbeitman}, \citenamefont {Chew},
  \citenamefont {Bernevig},\ and\ \citenamefont {T\"orm\"a}}]{Huhtinen2022}%
  \BibitemOpen
  \bibfield  {author} {\bibinfo {author} {\bibfnamefont {K.-E.}\ \bibnamefont
  {Huhtinen}}, \bibinfo {author} {\bibfnamefont {J.}~\bibnamefont
  {Herzog-Arbeitman}}, \bibinfo {author} {\bibfnamefont {A.}~\bibnamefont
  {Chew}}, \bibinfo {author} {\bibfnamefont {B.~A.}\ \bibnamefont {Bernevig}},\
  and\ \bibinfo {author} {\bibfnamefont {P.}~\bibnamefont {T\"orm\"a}},\ }\href
  {https://doi.org/10.1103/PhysRevB.106.014518} {\bibfield  {journal} {\bibinfo
   {journal} {Phys. Rev. B}\ }\textbf {\bibinfo {volume} {106}},\ \bibinfo
  {pages} {014518} (\bibinfo {year} {2022})}\BibitemShut {NoStop}%
\end{thebibliography}%
\end{document}


\title{Supplemental Material for ``Unconventional gapless semiconductor in an extended martini lattice in covalent honeycomb materials"}
\author{Tomonari Mizoguchi}
\affiliation{Department of Physics, Graduate School of Pure and Applied Sciences, University of Tsukuba, Tsukuba, Ibaraki 305-8571, Japan}
\author{Yanlin Gao}
\affiliation{Department of Physics, Graduate School of Pure and Applied Sciences, University of Tsukuba, Tsukuba, Ibaraki 305-8571, Japan}
\author{Mina Maruyama}
\affiliation{Department of Physics, Graduate School of Pure and Applied Sciences, University of Tsukuba, Tsukuba, Ibaraki 305-8571, Japan}
\author{Yasuhiro Hatsugai}
\affiliation{Department of Physics, Graduate School of Pure and Applied Sciences, University of Tsukuba, Tsukuba, Ibaraki 305-8571, Japan}
\author{Susumu Okada}
\affiliation{Department of Physics, Graduate School of Pure and Applied Sciences, University of Tsukuba, Tsukuba, Ibaraki 305-8571, Japan}
\date{\today}
\maketitle

\section{Tight-binding Hamiltonian and its property}
The Hamiltonian matrix for Fig.~1(a) of the main text is given as
\begin{eqnarray}
 H_{\bm{k}} = 
 \begin{pmatrix}
 V & t_2  & t_2 e^{i\bm{k}\cdot \bm{a}_1} &  t_2 e^{i\bm{k}\cdot \bm{a}_2} \\
 t_2 & 0 & t_1 + t_1^\prime  e^{i\bm{k}\cdot \bm{a}_1} & t_1 + t_1^\prime  e^{i\bm{k}\cdot \bm{a}_2} \\
 t_2 e^{-i\bm{k}\cdot \bm{a}_1}  & t_1 + t_1^\prime  e^{-i\bm{k}\cdot \bm{a}_1} & 0 & t_1 + t_1^\prime e^{i \bm{k}\cdot (\bm{a}_2 -\bm{a}_1)} \\ 
 t_2 e^{-i\bm{k}\cdot \bm{a}_2} &  t_1 + t_1^\prime  e^{- i\bm{k}\cdot \bm{a}_2} &  t_1 + t_1^\prime e^{i \bm{k}\cdot (\bm{a}_1 -\bm{a}_2)}  & 0 \\
 \end{pmatrix}. 
 \label{eq:Ham_extendedmartini}
 \end{eqnarray}
 
We note that the eigenenergies of $H_{\bm{k}}$ do not depend on the sign of $t_2$.
To be concrete, $H_{\bm{k}}$ satisfies
\begin{eqnarray}
H_{\bm{k}}(t_1,t_1^\prime, - t_2,V)  = \gamma H_{\bm{k}}(t_1, t_1^\prime, t_2,V)  \gamma, \label{eq:sym}
\end{eqnarray}
where we explicitly write the hopping parameter dependence of the Hamiltonian, 
and  $\gamma = \mathrm{diag}\left( -1,1,1,1 \right)$ is the unitary matrix. 
Note that $\gamma$ satisfies $\gamma^\dagger = \gamma$. 
Equation (\ref{eq:sym}) indicates 
that $H_{\bm{k}}(t_1,t_1^\prime,- t_2,V)$ and $H_{\bm{k}}(t_1, t_1^\prime,t_2,V)$ are unitary equivalent to each other, thus
their eigenvalues are the same.
Hence, the band structure is determined by the signs of $t_1$ and $t_1^\prime$. 

\section{Additional numerical results for tight-binding model}
In this section, we present additional numerical results for the tight-binding model. 
\subsection{Conventional martini lattice}
\begin{figure*}[tb]
\begin{center}
\includegraphics[clip,width = 0.85\linewidth]{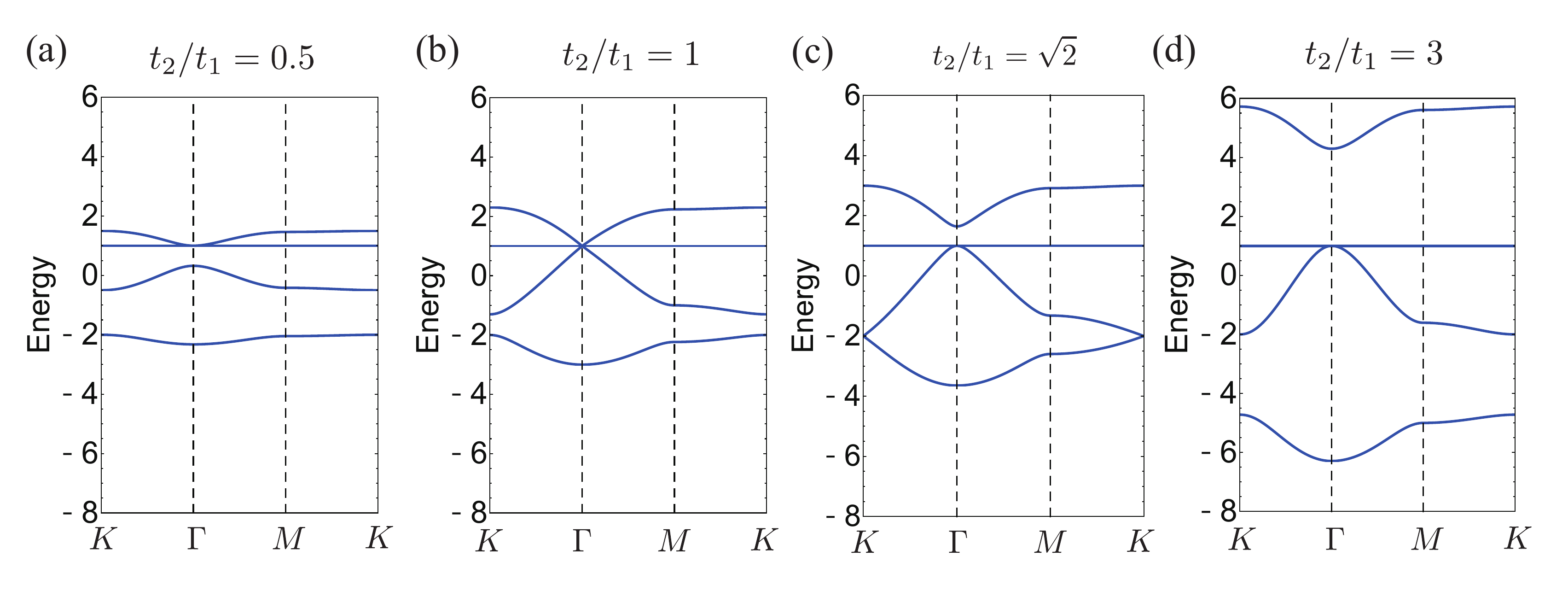}
\vspace{-10pt}
\caption{Band structures for $V= t_1^\prime = 0$. We set $t_1 = -1$.
The values of $t_2/t_1$ are indicated at the top of each panel. 
The high-symmetry momenta are $\Gamma = (0,0)$, $K = \left(\frac{4\pi}{3 a_0}, 0 \right)$, and $M = \left(\frac{\pi}{a_0}, \frac{\sqrt{3} \pi}{3a_0}\right)$.}
  \label{fig:band1}
 \end{center}
 \vspace{-10pt}
\end{figure*}
The model includes the conventional martini lattice model as a special case with $V= t_1^\prime = 0$.
Here we show the band structures of the conventional martini lattice model~\cite{Matsumoto2022}.
In Fig.~\ref{fig:band1}, we plot the band structures for several values of $t_2$. 
We see that the flat band and the quadratic band touching at 
$\Gamma$ point occurs for all case,
as we have argued in the main text. 
Similarly to the case we have discussed in the main text, the transition from the band insulator to 
the unconventional gapless semiconductor occurs at $t_2/t_1 =  1$;
at  $t_2/t_1 =  1$, the triple band touching occurs  
[Fig.~\ref{fig:band1}(b)].
The analytical derivation of this value is shown in Sec.~\ref{app:tbc}.
It is also worth noting that the Dirac cone formed by the first and the second lowest bands appears at $t_2/t_1 =  \sqrt{2}$ [Fig.~\ref{fig:band1}(c)].

\subsection{Effect of on-site potential}
\begin{figure*}[tb]
\begin{center}
\includegraphics[clip,width = 0.95\linewidth]{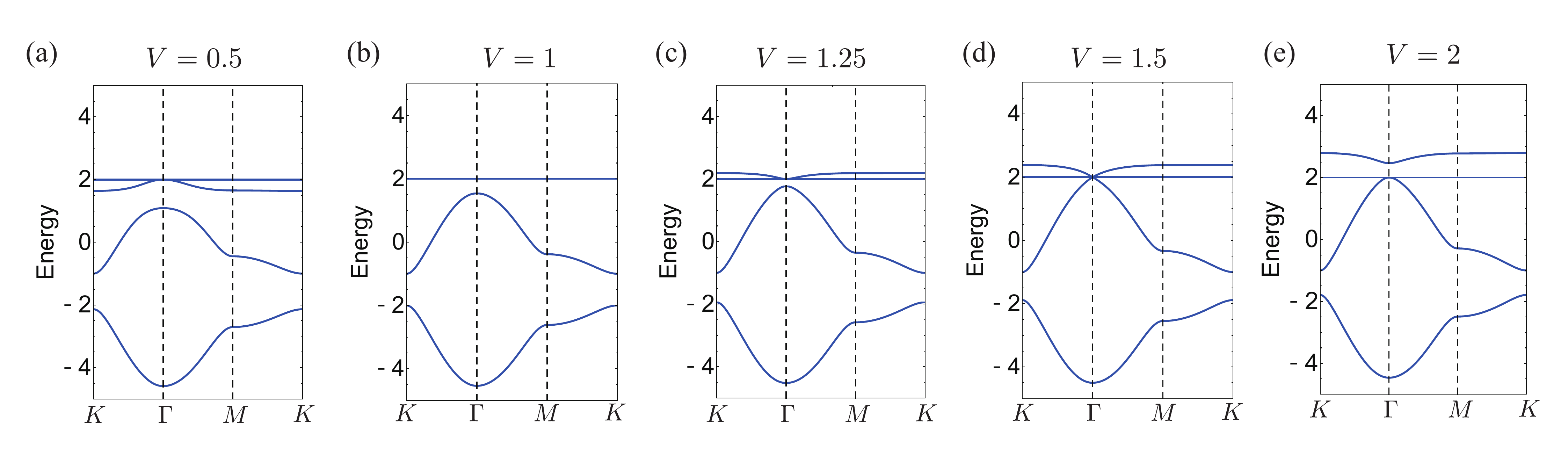}
\vspace{-10pt}
\caption{Band structures for $t_1 = t_1^\prime= t_2  = -1$ and finite $V$.
The values of $V$ are indicated at the top of each panel. }
  \label{fig:band2}
 \end{center}
 \vspace{-10pt}
\end{figure*}

Next, we investigate the effects of the on-site potential, $V$.
We first consider the case with $t_1 = t_1^\prime =t_2 = -1$. 
 The band structures for several values of $V$ are shown in Fig.~\ref{fig:band2}. 
In this case, the transition point between the band insulator and the unconventional gapless semiconductor
is at $V= 1.5$ [Fig.~\ref{fig:band2}(d)]. 
Besides, at $V= 1$, the doubly degenerate flat band appears [Fig.~\ref{fig:band2}(b)], similarly to Fig.~2(b) in the main text.
The analytic derivation of this value is presented in Appendix~\ref{app:ddfb}.

\begin{figure*}[tb]
\begin{center}
\includegraphics[clip,width = 0.95\linewidth]{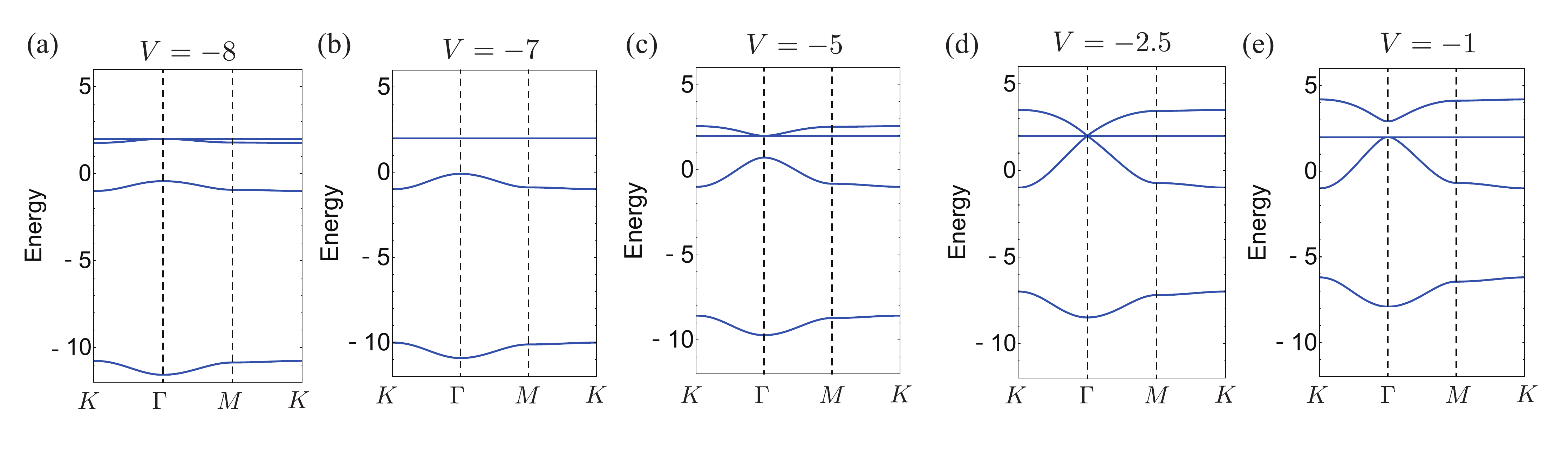}
\vspace{-10pt}
\caption{Band structures for $t_1 = t_1^\prime = -1$, $t_2 = -3$ and finite $V$. 
The values of $V$ are indicated at the top of each panel.}
  \label{fig:band3}
 \end{center}
 \vspace{-10pt}
\end{figure*}

We also consider the case with $t_1 = t_1^\prime  = -1$ and $t_2 = -3$.
In this case, the transition from the band insulator to the conventional semiconductor occurs at $V = -2.5$ [Fig.~\ref{fig:band3}(d)] ,
whereas the doubly degenerate flat band occurs at $V = -7$  [Fig.~\ref{fig:band3}(b)]. 

\section{Analytic derivations of characteristic band structures}
In this section, 
we analytically investigate the origins of the characteristic band structures,
namely the triple band touching at $\Gamma$ point and the doubly degenerate flat band.

\subsection{Triple band touching at $\Gamma$ point \label{app:tbc}}
We first examine the condition of the triple band touching 
by explicitly solving the eigenvalue equation at $\Gamma$ point with a certain \textit{ansatz}.
Specifically, as we have seen, two out of four eigenstates have the eigenvalue $\varepsilon^{\rm BF} =-(t_1 + t_1^\prime)$ at $\Gamma$ point, 
so we investigate the remaining eigenvalues.
Then, recalling that $\bm{\varphi}_{\bm{k}= \Gamma} $ is orthogonal to $(0,1,1,1)^{\rm T}$, 
 we assume the following form of the eigenvector (up to the normalization constant):
\begin{eqnarray}
\bm{\varphi}^\prime_{\bm{k} = \Gamma} = 
\begin{pmatrix}
\alpha \\
1 \\
1\\
1\\
\end{pmatrix}. \label{eq:ansatz}
\end{eqnarray} 
Acting the Hamiltonian to $\bm{\varphi}^\prime_{\bm{k}}$, we have
 \begin{eqnarray}
H_{\Gamma} \bm{\varphi}^\prime_{\Gamma} 
&=& \begin{pmatrix}
V & t_2 & t_2 & t_2 \\
t_2 & 0 & t_1+t_1^\prime & t_1+ t_1^\prime  \\
t_2 & t_1+t_1^\prime & 0 & t_1+t_1^\prime \\
t_2 & t_1+& t_1+t_1^\prime & 0 \\
\end{pmatrix}
\begin{pmatrix}
\alpha \\
1 \\
1\\
1\\
\end{pmatrix} \nonumber \\
 &=& \begin{pmatrix}
\alpha V + 3 t_2 \\
 \alpha t_2+ 2(t_1+t_1^\prime) \\
 \alpha t_2+ 2(t_1+t_1^\prime) \\
 \alpha t_2+2( t_1+t_1^\prime) \\
 \end{pmatrix}. \nonumber \\
\end{eqnarray}
In order that the triple band touching occurs, it is sufficient that $\bm{\varphi}^\prime_{\Gamma}$
is the eigenvector of $H_{\Gamma} $ with the eigenvalue $ -(t_1+t_1^\prime)$.
If it is the case, the following conditions have to be satisfied:
\begin{eqnarray}
\alpha V + 3 t_2 = -\alpha (t_1+t_1^\prime), \label{eq:cond1}
\end{eqnarray}
and
\begin{eqnarray}
  \alpha t_2+ 2(t_1+ t_1^\prime) = - (t_1+t_1^\prime). \label{eq:cond2}
\end{eqnarray}
From Eq.~(\ref{eq:cond1}), we have $\alpha = -\frac{3t_2}{V+t_1+t_1^\prime}$.
Similarly, from Eq.~(\ref{eq:cond2}), we have $\alpha = -\frac{3(t_1+t_1^\prime)}{t_2}$.
Then, we can conclude that 
the triple band touching occurs for the fine-tuned parameters where the above two conditions
are satisfied simultaneously. 
Specifically, the condition for the tight-binding parameters is
\begin{eqnarray}
t_2^2 = (t_1+ t_1^\prime) (V +t_1+ t_1^\prime). \label{eq:cond3}
\end{eqnarray}
The parameter sets for Figs.~\ref{fig:band1}(b), \ref{fig:band2}(d), and \ref{fig:band3}(d), as well as Fig.~2(d) in the main text,
indeed satisfy Eq.~(\ref{eq:cond3}).

\subsection{Doubly degenerate flat band \label{app:ddfb}}
Next, we elucidate the origin of doubly degenerate flat bands.
Generically, the doubly degenerate flat band at $\varepsilon^{\rm FB} =-( t_1 + t_1^\prime) $ appears when the Hamiltonian can be written as
\begin{eqnarray}
H_{\bm{k}} = \Psi^\prime_{\bm{k}} h_{\bm{k}}^\prime \Psi^{\prime \dagger}_{\bm{k}} - (t_1 + t_1^\prime)I_4, \label{eq:Ham_MOrep_double}
\end{eqnarray}
where $\Psi_{\bm{k}}^\prime$ is a $4 \times 2$ matrix and $h^\prime_{\bm{k}}$ is a $2 \times 2$ Hermitian matrix.

Here, we make assumptions on the forms of $h^\prime_{\bm{k}}$ and $ \Psi^\prime_{\bm{k}}$.
To be specific, we assume 
\begin{eqnarray}
h_{\bm{k}}^\prime = \begin{pmatrix}
t_1 & 0 \\ 0 & t_1^\prime \\
\end{pmatrix}, \label{eq:hp}
\end{eqnarray}
and 
\begin{eqnarray}
\Psi^\prime = 
\begin{pmatrix}
0 & \beta \\
1 & 1 \\
1 & e^{-i\bm{k}\cdot \bm{a}_1} \\
1 & e^{-i\bm{k}\cdot \bm{a}_2} \\
\end{pmatrix}, \label{eq:psip}
\end{eqnarray}
where $\beta$ is the variable to be determined.
Substituting Eqs.~(\ref{eq:hp}) and (\ref{eq:psip}) into Eq.~(\ref{eq:Ham_MOrep_double}), we have 
\begin{eqnarray}
H_{\bm{k}} = 
\begin{pmatrix}
\beta^2 t_1^\prime - (t_1 + t_1^\prime) & \beta t_1^\prime &   \beta t_1^\prime e^{i\bm{k}\cdot \bm{a}_1}  &  \beta t_1^\prime e^{i\bm{k}\cdot \bm{a}_2} \\
\beta t_1^\prime & 0 & t_1 + t_1^\prime  e^{i\bm{k}\cdot \bm{a}_1} & t_1 + t_1^\prime  e^{i\bm{k}\cdot \bm{a}_2} \\
\beta t_1^\prime e^{-i\bm{k}\cdot \bm{a}_1}  & t_1 + t_1^\prime  e^{-i\bm{k}\cdot \bm{a}_1} & 0 & t_1 + t_1^\prime e^{i \bm{k}\cdot (\bm{a}_2 -\bm{a}_1)} \\ 
\beta t_1^\prime e^{-i\bm{k}\cdot \bm{a}_2} &  t_1 + t_1^\prime  e^{- i\bm{k}\cdot \bm{a}_2} &  t_1 + t_1^\prime e^{i \bm{k}\cdot (\bm{a}_1 -\bm{a}_2)}  & 0 \\
\end{pmatrix}. \label{eq:MOrep_2}
\end{eqnarray}
Comparing Eq.~(\ref{eq:Ham_extendedmartini}) with Eq.~(\ref{eq:MOrep_2}), 
we find that $\beta$ has to satisfy
\begin{eqnarray}
 \beta = \frac{t_2}{t_1^\prime}. \label{eq:beta}
\end{eqnarray}
Substituting this into $(1,1)$-component of Eq.~(\ref{eq:MOrep_2}) and comparing it with that of Eq.~(\ref{eq:Ham_extendedmartini}) we have
\begin{eqnarray}
V = \frac{t_2^2}{t_1^\prime}- (t_1 + t_1^\prime).  \label{eq:cond_dfb}
\end{eqnarray}
Therefore, the doubly degenerate flat band occurs for the parameter sets satisfying Eq.~(\ref{eq:cond_dfb}).
The parameter sets for Figs.~\ref{fig:band2}(b) and \ref{fig:band3}(b), as well as Fig.~2(b) in the main text, 
indeed satisfy Eq.~(\ref{eq:cond_dfb}).
It is also worth pointing out that, for the conventional martini lattice (i.e., $t_1^\prime = 0$), 
the doubly degenerate flat band occurs only in the limit of $V\rightarrow \infty$
since Eq.~(\ref{eq:beta}) requires $\beta \rightarrow \infty$. 

 \section{DFT calculations}
We perform the density functional theory calculation by using STATE program package~\cite{state1,state3}. 
Exchange-correlation potential among interacting electrons is treated by generalized gradient approximation with the Perdew--Burke--Ernzerhof (PBE) functional~\cite{gga}. An ultrasoft pseudopotential is adopted to described the interaction between valence electrons and ions~\cite{ppot}. Valence wave functions and the deficit charge density were expanded in terms of plane-wave basis sets with cutoff energies of 25 and 225 Ry, respectively.  
A Brillouin-zone integration was carried out with a 1$\times$10$\times$1 $\bm{k}$-mesh for the self-consistent electronic structure calculations. Atomic coordinates of martini structures are optimized till the force less than 1.33$\times$10$^{-3}$ hartree.
\section{Partially hydrogenated silicene}
\begin{figure*}[htb]
\begin{center}
\includegraphics[width = 0.5\hsize]{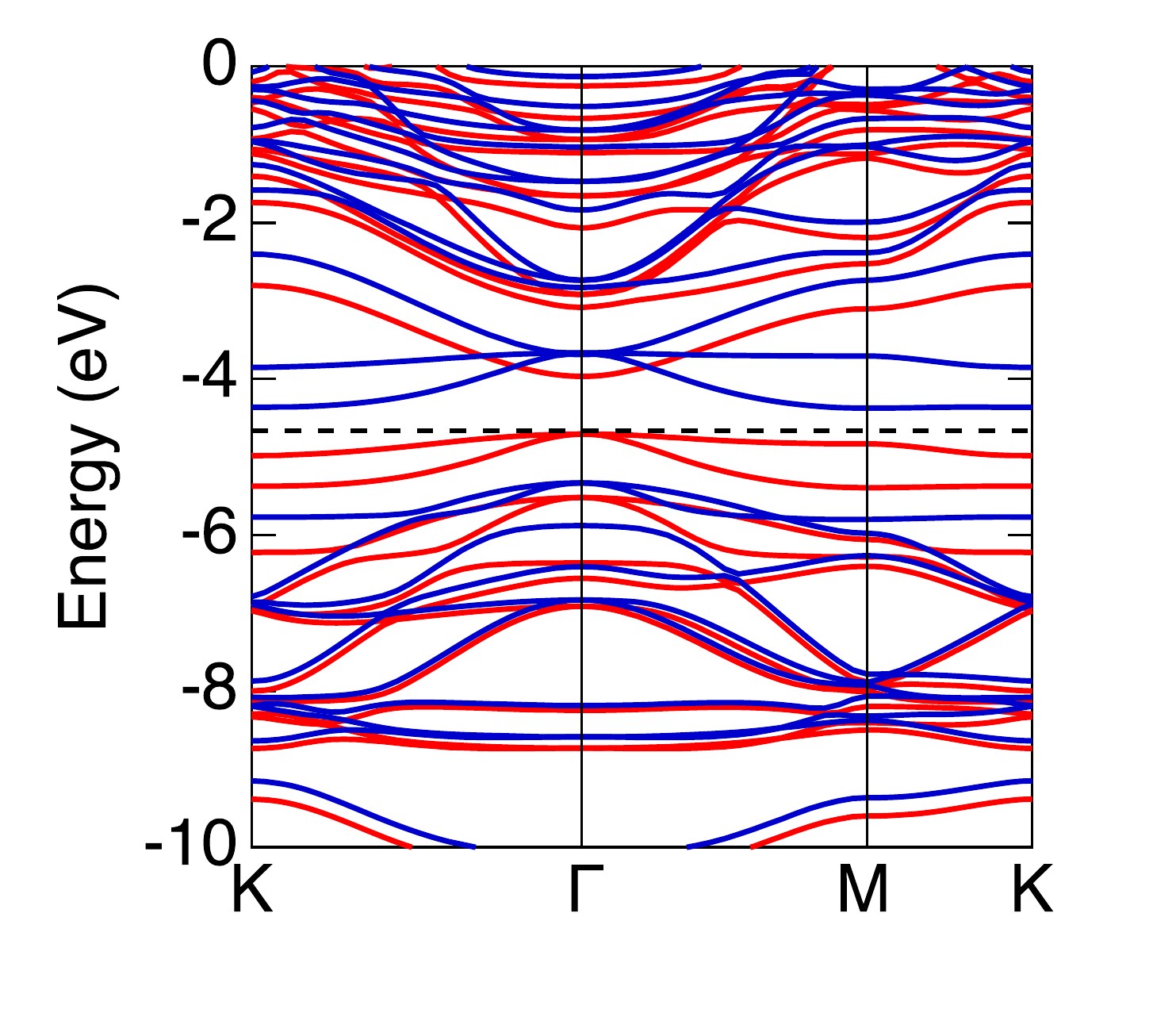}
\caption{
Band structure of partially hydrogenated silicene.  
Red and blue curves denote the energy band for majority and minority states, respectively. Energies are measured from the vacuum level. The horizontal dotted line indicates the Fermi level.
}
  \label{fig:band_sih}
 \end{center}
\end{figure*}

The partially hydrogenated silicene has a type-I flat band arrangement 
as the case of fluorinated graphene (Fig.~\ref{fig:band_sih}).
Owing to the partial occupation of the flat band, the hydrogenated silicene exhibits spin polarization 
with the magnetic moment of 0.15 $\mu_B$/{\AA}$^2$, 
where the polarized electron spin has the same spatial distribution as the flat-band wave function of the extended martini model
at $\Gamma$ and the Fermi level.

\section{BCN heterostructures obtained by B/N substitution}
\begin{figure*}[htb]
\begin{center}
\includegraphics[width = 0.75\hsize]{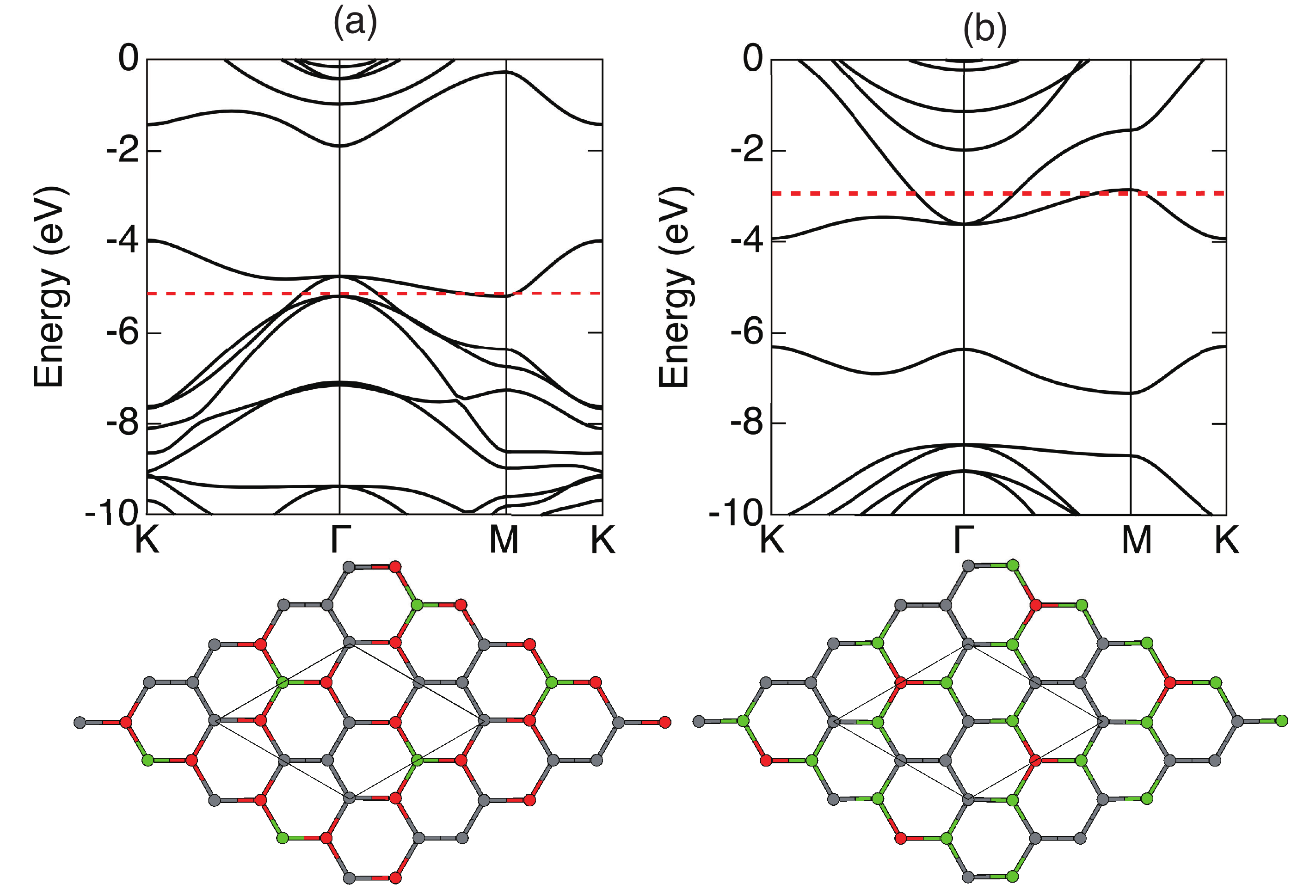}
\caption{
Electronic and geometric structures of BCN heterostructures with (a) C-B and (b) C-N borders. Energies are measured from the vacuum level. Red dotted horizontal lines denote the Fermi level. Red, gray, and green balls denote B, C, and N atoms, respectively. The rhombus denotes the unit cell of BCN heterostructures.}
  \label{fig:band_bnc}
 \end{center}
\end{figure*}

In addition to the adsorption, 
the chemical substitution is another method of controlling the structure of the $\pi$-electron network.
Here we show the results of graphene with B/N substitution.
The electronic structure of BCN heterostructures with C-B and C-N borders are shown in FIg.~\ref{fig:band_bnc}.
The heterostructures are obtained by substituting four of eight C atomic sites in 2$\times$2 unit cell of graphene by B an N atoms.
Both heterostructures are metals in which the Fermi level crosses parabolic and flat dispersion bands owing to the substantial band width of the flat bands.
The heterostructures with C-B and C-N borders have type-I and type-II flat band arrangements, respectively.
Therefore, a sign of electron transfer $t_1$ is tunable by controlling the border atomic arrangement. 
Furthermore, the band filling of the flat band may also be tunable by the atomic species 
which are substitutionally doped in graphene. 
